\documentclass[pra,twocolumn, showpacs, showkeys, secnumarabic, aps, amsmath, amssymb, nofootinbib, superscriptaddress, longbibliography, floatfix, table-of-contents, eqsecnum, dblfloatfix]{revtex4-2}

\usepackage[pdftex]{graphicx}
\usepackage{mathrsfs}
\usepackage[colorlinks, breaklinks, urlcolor={blue}, linkcolor={red}, citecolor={blue}]{hyperref}
\usepackage{array}
\usepackage{amsmath}
\usepackage{type1cm}
\usepackage{lettrine}
\usepackage[english]{babel}
\usepackage{lmodern}
\usepackage{microtype}
\usepackage{booktabs}
\usepackage[T1]{fontenc}
\usepackage[boxed, vlined]{algorithm2e}
\usepackage{caption}
\usepackage{braket}
\usepackage{xcolor}

\def\nn{\nonumber}
\begin{document}

\title{Accessible and inaccessible quantum coherence in relativistic quantum systems}

\author{Saveetha Harikrishnan}
\email[]{saveethah@citchennai.net}
\affiliation{Centre for Quantum Science \& Technology, Chennai Institute of Technology, Chennai 600069, India}

\author{Segar Jambulingam}
\email[]{segar@rkmvc.ac.in}
\affiliation{Department of Physics, Ramakrishna Mission Vivekananda College, Mylapore, Chennai 600 004, India}

\author{Peter P. Rohde}
\email[]{dr.rohde@gmail.com}
\homepage[]{https://www.peterrohde.org}
\affiliation{Centre for Quantum Software \& Information (UTS:QSI), University of Technology Sydney, Ultimo, NSW 2007, Australia}
\affiliation{Hearne Institute for Theoretical Physics, Department of Physics \& Astronomy, Louisiana State University, Baton Rouge LA, United States}

\author{Chandrashekar Radhakrishnan}
\email[]{chandrashekar10@gmail.com}
\affiliation{Centre for Quantum Information, Communication and Computing, Indian Institute of Technology Madras, Chennai 600036, India}

\affiliation{Centre for Quantum Science \& Technology, Chennai Institute of Technology, Chennai 600069, India}
\homepage[]{https://sites.google.com/view/chandrashekar}

\begin{abstract}
The quantum coherence of a multipartite system is investigated when some of the parties are moving with uniform acceleration and the analysis is 
carried out using the single mode approximation. Due to acceleration the quantum coherence is divided into two parts as accessible and 
inaccessible coherence and the entire analysis has been carried out in the single-mode approximation. First we investigate tripartite systems, 
considering both GHZ and W-states.  We find that the quantum coherence of these states does not vanish in the limit of infinite acceleration, 
rather asymptoting to a non-zero value. These results hold for both single- and two-qubit acceleration. In the GHZ and W-states the coherence 
is distributed as correlations between the qubits and is known as global coherence. But quantum coherence can also exist due to the superposition 
within a qubit, the local coherence.  To study the properties of local coherence we investigate separable state. The GHZ state, W-state and 
separable states contain only one type of coherence. Next we consider the $W \bar{W}$ and star states in which both local and global coherences coexist. 
We find that under uniform acceleration both local and global coherence show similar qualitative behaviour. Finally we derive analytic expressions for 
the quantum coherence of $N$-partite GHZ and W-states for $n<N$ accelerating qubits. We find that the quantum coherence of a multipartite GHZ state 
falls exponentially with the number of accelerated qubits, whereas for multipartite W-states the quantum coherence decreases only polynomially. 
We conclude that W-states are more robust to Unruh decoherence and discuss some potential applications in satellite-based quantum communication 
and black hole physics. 
\end{abstract}
\maketitle


\section{Introduction}

Entanglement is a widely studied quantum resource with applications in quantum computing \cite{shor1998quantum}, quantum algorithms \cite{deutsch1992rapid, 
grover1996fast, shor1999polynomial}, metrology \cite{giovannetti2006quantum}, teleportation \cite{bennett1993teleporting, bouwmeester1997experimental}, and 
cryptography \cite{bennett1984quantum, bennett1992experimental}. In recent times, it has become more apparent that there are several other quantum properties 
such as non-locality \cite{bancal2009quantifying}, steering \cite{skrzypczyk2014quantifying}, discord \cite{ollivier2001quantum} and coherence 
\cite{baumgratz2014quantifying}, which can also be used as a resource. Based on their relative presence, it is known that there is a hierarchy among these 
different resources. Of all these resources, quantum coherence is more extensively present, even when the quantum system does not have steering, entanglement 
or discord. Hence an investigation of quantum coherence can provide more complete information about the `quantumness' of physical systems. Consequently, 
several studies have been performed into quantum coherence with the aim to understand the fundamental quantum behaviour of systems. Some important 
investigations carried out so far are on the measurement of quantum coherence \cite{baumgratz2014quantifying, radhakrishnan2016distribution}, the formulation 
of resource theory of quantum coherence \cite{winter2016operational, brandao2015reversible, chitambar2016critical, chitambar2015relating}, 
the effect of external environments on coherence \cite{chanda2016delineating, radhakrishnan2019dynamics}
and finally applications of quantum coherence \cite{hillery2016coherence, zhang2019demonstrating, ma2019operational}.  

Investigations in quantum information theory are generally carried out in inertial reference frames. An extension to non-inertial reference frames has been 
discussed through several works \cite{peres2004quantum, dunningham2009entanglement, mann2012relativistic, alsing2002lorentz}. In this context entanglement in 
non-inertial frames of reference has been an important area of research \cite{alsing2012observer, li2003relativistic, shi2004entanglement}.
Initial works \cite{fuentes2005alice} in this direction considered a family of peaked Minkowski wave packets which admits only a single 
Unruh mode. Later on the entanglement of a general set of states has been discussed in a multimode setting \cite{Fuentes-beyondsinglemodeapprx, Bruschi2013}.
Though the multimode approximation is more generally valid, several studies are still carried out in the single mode approximation \cite{Dong_singlemodeapprox_WClass_2020}, due to the ease of obtaining a clear analytic expression. Another approach is also available in non-inertial scenario by considering Unruh-Dewitt detector model\cite{Unruh1976, Quan_comm2013ralph}.  
Several interesting results have been obtained in the fields of black hole physics \cite{fuentes2005alice, kabat1995black, henderson2018harvesting, 
martin2010unveiling, martin2010quantum, terashima2000entanglement}, quantum error correction \cite{verlinde2013black}, relativistic quantum metrology 
\cite{ahmadi2014relativistic, ahmadi2014quantum, tian2015relativistic, du2021fisher}, relativistic teleportation \cite{friis2013relativistic, lin2015quantum, 
grochowski2017effect, koga2018quantum} and communication \cite{downes2013quantum, landulfo2016nonperturbative, kravtsov2018relativistic, 
radchenko2014relativistic}. Similar studies on relativistic effects on quantum discord \cite{datta2009quantum, jung2013quantum} have also been performed. 
Recent experimental advances have increased the scope of quantum technologies from applications in terrestrial situations \cite{mattle1996dense, 
ursin2007entanglement, jin2010experimental} to satellite-based space level technologies \cite{brito2021satellite, liao2017satellite, yin2012quantum, 
villar2020entanglement, yin2017satellite, yin2020entanglement, yin2017satellite2}. Hence, we investigate quantum coherence under uniform acceleration since 
compared to entanglement or other resources it is more commonly found and can be measured more easily.  

The effect of relativistic motion on quantum coherence has been studied in the following works \cite{wang2016irreversible, huang2018quantum, 
he2018multipartite, zeng2021distribution}. In Ref.~\cite{wang2016irreversible}, the relativistic effects on the quantum coherence between a pair of 
Unruh-Dewitt detectors is studied. Here they find that compared to entanglement, quantum coherence is more robust to Unruh decoherence. A generalisation of 
this result to tripartite system was performed by \cite{he2018multipartite}. Both these works consider a massless scalar field. The relativistic coherence of a system of Dirac fields was considered in \cite{zeng2021distribution}. In our work we consider the modes of a massless scalar field and measure the coherence 
between different modes as well as the coherence within a mode.  Here we consider a uniformly accelerating system which can be described
using a family of peaked Minkowski wave packets and consequently we work in the single mode setting. We estimate the 
information-theoretic change in the coherence of the system using the $\ell_{1}$-norm of coherence. For a complete study of tripartite systems, we investigate 
both the SLOCC (Stochastic Local Operations \& Classical Communication) class of states, namely the GHZ (Greenberger-Horne-Zeilinger) and W class of states. 
But these two states have only global coherence which arises due to correlation between qubits. For completeness sake, we investigate the relativistic effects 
on separable states with only local coherence, a type of quantum coherence arising due to superposition within qubits. We also look into two different 
tripartite states, namely the $W \bar{W}$ and star states in which both global and local coherence coexist.  

The structure of the manuscript is as follows: In Sec.~\ref{sec:rel_and_q_coh} we discuss the notion of relativity in the field of quantum information and also introduce the $\ell_{1}$-norm of coherence which is used to measure coherence in our work. The SLOCC class of states is studied in detail in 
Sec.~\ref{sec:rel_eff_slocc} with the GHZ class and W class forming two different subsections. The separable state and an introduction to the notions of local 
coherence and global coherence is described in Sec.~\ref{sec:separable}. The tripartite systems with both local and global coherence are analysed in 
Sec.~\ref{sec:tripartite}. In Sec.~\ref{sec:apps} we discuss certain applications related to satellite-based communication as well as black holes. Finally in 
Sec.~\ref{sec:discussion} we discuss our results.  

\section{Relativity \& quantum coherence} \label{sec:rel_and_q_coh}

\begin{figure}
\includegraphics[width=0.99\columnwidth]{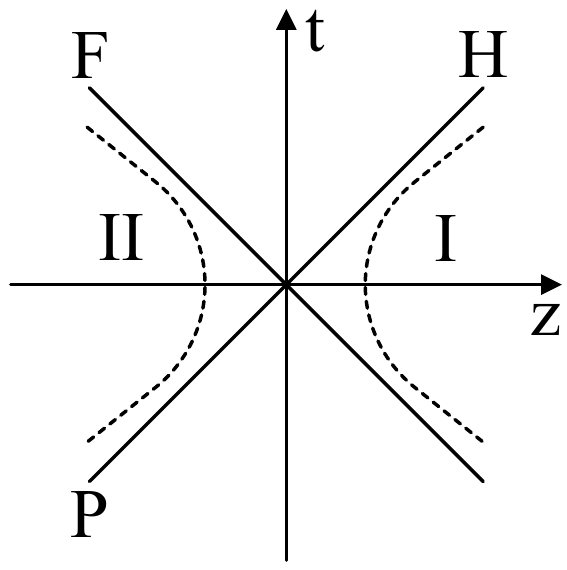} 
\caption{The $(z,t)$ plane of the Minkowski co-ordinates is divided into four regions of the Rindler co-ordinates. The solid lines are the future ($F$) and 
past ($P$) event horizons and the dashed lines are the trajectories of the uniformly accelerated observers. Here $H$ represents the horizon.}
\label{fig1}
\end{figure}

To describe events independent of the inertial frame of reference, the combination of three space and one time dimension known as Minkowski space are used. For non-inertial reference frames we need to use the Rindler co-ordinates. The Minkowski and Rindler co-ordinates are related to each other via a Rindler 
transformation. For a situation with one spatial and one time dimension $(z,t)$,  the Minkowski space-time is divided into four wedges as shown in 
Fig.~\ref{fig1}. The regions $F$ and $P$ represent the future and the past light cones and the regions I and II are the two causally disconnected Rindler 
regions. The equations $|z|=t$ and $|z|=-t$ describe the future and past event horizons.  For situations with one spatial and one time dimension $(z,t)$, the 
world lines of uniformly accelerated observers in Minkowski co-ordinates correspond to a hyperbola. The two branches of the hyperbola constitute the regions I 
and II of the Rindler co-ordinates.  
The co-ordinates of the two regions are,
\begin{align}
t = a^{-1} e^{a \xi} \sinh a \tau,         z=a^{-1} e^{a \xi} \cosh a \tau; |z|<t,   \nonumber \\
t = -a^{-1} e^{a \xi} \sinh a \tau,      z=a^{-1} e^{a \xi} \cosh a \tau; |z|>t.
\end{align}
where $\xi$ is the space-like co-ordinate, $\tau$ is proper time, and $a$ is acceleration.  

Alice, Bob and Charlie each having monochromatic detector with their corresponding frequencies, $\omega_{1}$, $\omega_{2}$ and $\omega_{3}$ of a free massless scalar field in a 
Minkowski space time. A maximally entangled GHZ state in the framework reads,
\begin{align}
    \frac{|0_{\omega_{1}} \rangle^{\mathcal{M}} |0_{\omega_{2}} \rangle^{\mathcal{M}} |0_{\omega_{3}} \rangle^{\mathcal{M}} + |1_{\omega_{1}} \rangle^{\mathcal{M}} |1_{\omega_{2}} \rangle^{\mathcal{M}} |1_{\omega_{3}} \rangle^{\mathcal{M}}}{\sqrt{2}}, \label{GHZdefinition}
\end{align}
where $|0_{\omega_{i}} \rangle^{\mathcal{M}}$ ($|1_{\omega_{i}} \rangle^{\mathcal{M}}$) is the vacuum (single particle excitation) state of frequency 
$\omega_{i}$ in Minkowski space. Initially when all three parties are in inertial frames then $|0_{\omega_{1}} \rangle^{\mathcal{M}}$, 
$|0_{\omega_{2}} \rangle^{\mathcal{M}}$ and $|0_{\omega_{3}} \rangle^{\mathcal{M}}$ are their respective ground states. Their excited states can be obtained by applying their corresponding Minkowski creation operators as follows,

\begin{align}
|n_{\omega_{i}} \rangle^{\mathcal{M}} &= \frac{(\hat{a}_{\omega_{i}}^{\dag})^{n}}{\sqrt{n !}} |0_{\omega_{i}} \rangle^{\mathcal{M}}.
\end{align}


If Charlie starts moving with uniform acceleration the wavefunction becomes higly delocalized in space and the quantum state corresponding to the frequency $\omega_{3}$ can be specified using Rindler or Unruh co-ordinates \cite{Fuentes-beyondsinglemodeapprx}. In both the Rindler and Unruh basis, the initial Minkowski space is divided into two regions which are casually disconnected from each other. Let $\Omega$ be the dimensionless Rindler frequency and the Unruh modes are also sharply peaked at the same frequency. Using an analytic continuation arguement it was shown \cite{Fuentes-beyondsinglemodeapprx} that Unruh mode is a purely positive frequency linear combinations of the Minkowski modes. But the definition of positive frequency in the Rindler mode differs from the positive frequency definition on the Minkowski mode. So we can relate the Minkowski and the Rindler modes via the Unruh modes. The field in each of these three bases are expanded as follows:
\begin{align}
\phi &=  \int_{0}^{\infty} (a_{\omega,\mathcal{M}} u_{\omega,\mathcal{M}} + a^\dagger_{\omega,\mathcal{M}}u^*_{ \omega,\mathcal{M}}) \,d\omega \nonumber \\
     &=  \int_{0}^{\infty} (A_{\Omega,R} u_{\Omega,R} + A^\dagger_{\Omega,R} u^*_{\Omega,R} \nonumber \\
     &  +   A_{\Omega,L} u_{\Omega,L} + A^\dagger_{\Omega,L} u^*_{\Omega,L}) \,d\Omega  \nonumber \\
     &=  \int_{0}^{\infty} (b_{\Omega,I} u_{\Omega,I} + b^\dagger_{\Omega,I} u^*_{\Omega,I} \nonumber \\
     &  +   b_{\Omega,II} u_{\Omega,II} + b^\dagger_{\Omega,II} u^*_{\Omega,II}) \,d\Omega ,  \nonumber \\
\end{align}
where $a_{\omega,\mathcal{M}}$ is the Minkowski annihilation operator and $A_{\Omega,R} \& A_{\Omega,L}$ are the Unruh annihilation operator 
for the right and left regions and $b_{\Omega,I} \& b_{\Omega,II}$ are the Rindler annihilation operators in $I \& II$ regions. 
The operators obey the bosonic commutation relations $[a_{\omega_1,\mathcal{M}} ,a^\dagger_{\omega_2,\mathcal{M}}] = \delta_{\omega_1 \omega_2}$, $[A_{\Omega_1,R}, A^\dagger_{\Omega_2,R}]$ =  $[A_{\Omega_1,L}, A^\dagger_{\Omega_2,L}]= \delta_{\Omega_1 \Omega_2}$
and $[b_{\Omega_1,I}, b^\dagger_{\Omega_2,I}] = [b_{\Omega_1,II}, b^\dagger_{\Omega_2,II}] = \delta_{\Omega_1 \Omega_2}$.
For the Unruh mode, the commutator between operators in the R and L region vanish. Similarly the commutator between the operators 
in region $I$ and $II$ vanish. The creation and annihilation operators of the Minkowski and Unruh bases do not mix and hence we 
have $|0\rangle_M = |0\rangle_U= \prod_{\Omega}|0{_\Omega}\rangle_U$. But the state $|0\rangle_U$ does not coincide with the Rindler 
vacuum and we have
\begin{align}
|0_\Omega\rangle_U = \sum_N \frac{\tanh^{n} r_{\Omega}}{\cosh {r}_{\Omega}} |n_\Omega\rangle_I |n_\Omega\rangle_{II}.     
\end{align}
Here $|n_{\Omega}\rangle_{I}$ is the $n^{th}$ excited state of the Rindler $I$ vacuum state $|0_{\Omega}\rangle_{I}$.

We consider a wave packet which is narrowly peaked in $\Omega$ and for this the Unruh and Rindler commutators read $[A_{\Omega,R}, A^\dagger_{\Omega,R}]$ =  $[A_{\Omega,L}, A^\dagger_{\Omega,L}]= 1 $
and $[b_{\Omega,I}, b^\dagger_{\Omega,I}] = [b_{\Omega,II}, b^\dagger_{\Omega,II}] = 1$. Under these ideal conditions, the most general creation operator of a purely positive Minkowski frequency can be written as,

\begin{align}
a^\dagger_{\Omega,U} = q_L A^\dagger_{\Omega,L} + q_R A^\dagger_{\Omega,R}.    
\end{align}
Here the factors $q_R$ and $q_L$ are complex numbers with $|q_R|^2 + |q_L|^2 = 1$. Under these conditions we have

\begin{align}
 a^\dagger_{\Omega,U} |0_\Omega \rangle_U = \sum_{n=0}^{\infty}\frac{\tanh^n r_\Omega}{\cosh r_\Omega} (\frac{\sqrt{n+1}}{\cosh r_\Omega}) |\Phi^n_\Omega \rangle , \nonumber \\
 |\Phi^n_\Omega \rangle = q_L |n_\Omega \rangle_I |(n+1)_\Omega\rangle_{II} + q_R|(n+1)_\Omega\rangle_I|n_\Omega \rangle_{II},
\end{align}
where, in general we consider $q_R=1$ and $q_L=0$. Let us consider a Minkowski smearing function which is a Gaussian in $\ln(\omega l)$,
\begin{align}
f(\omega) = \left( \frac{\lambda}{\pi \omega^2} \right)^{1/4} \exp \Big\{ \frac{-1}{2} \lambda [\ln(\omega/\omega_0)]^2 \Big\} (\omega/\omega_0)^{-i\mu},    
\end{align}
When the uniformly accelerated particle has the above smearing function and has negligible overlap with the other states, then it is well approximated by a single Unruh frequency. Thus we use this monochromatic wave approximation in our investigation. Under this 
condition, the Minkowski and Rindler modes can be connected via the relations: 

%
\begin{align}
    \hat{a}_{\omega_{3}}^{\dag} &= \hat{b}^{\dag}_{\Omega_{3}I} \cosh r  - \hat{b}_{\Omega_{3}II}  \sinh r 
    = \hat{S}_{\Omega_{3}} \hat{b}^{\dag}_{\Omega_{3}I} \hat{S}^{\dag}_{\Omega_{3}} \nonumber \\
    \hat{a}_{\omega_{3}} &= \hat{b}_{\Omega_{3}I} \cosh r - \hat{b}^{\dag}_{\Omega_{3}II}  \sinh r 
    = \hat{S}_{\Omega_{3}} \hat{b}_{\Omega_{3}I} \hat{S}^{\dag}_{\Omega_{3}}, 
\end{align}
where,
\begin{align}
    \hat{S}_{\Omega}(r) = \exp[r (\hat{b}^{\dag}_{\Omega I} \hat{b}^{\dag}_{\Omega II} - \hat{b}_{\Omega I} \hat{b}_{\Omega II})].
\end{align}
Here the operator $(\hat{S})$ effecting the transformation from Minkowski co-ordinates to Rindler co-ordinates is structurally identical to the two mode squeezing operator,
\begin{align}
    \hat{S}(\zeta) =\exp(\zeta^{*} ab - \zeta a^{\dag} b^{\dag}),
\end{align}
in the quantum optics context. Hence for a non-inertial observer, the single mode Minkowski vacuum becomes a two mode squeezed state in the Rindler vaccum,
\begin{align}
|0_{\omega} \rangle^{\mathcal{M}} &=  \hat{S}_{\Omega_{3}}(r) ( | 0 \rangle_{I} \otimes | 0 \rangle_{II} )  \nonumber \\
&= \frac{1}{\cosh r} \sum_{n=0}^{\infty} \tanh^{n} r |n_{\Omega} \rangle_{I} |n_{\Omega} \rangle_{II},
\end{align}
where $\cosh r = (1-e^{-2 \pi \Omega})^{-1/2}$ and $\Omega = |\omega| c/a$. The factors $\omega$ and $c$ are the wave vector and 
velocity of light respectively. Here $|n_{\Omega} \rangle_{I}$ and $|n_{\Omega} \rangle_{II}$ are the mode decompositions in Rindler regions I and region II respectively. For the single particle excitation state we have,
\begin{align}
    |1_{\omega} \rangle^{\mathcal{M}} &= \hat{a}_{\omega_{3}}^{\dag}  |0_{\omega_{3}} \rangle ^{\mathcal{M}} = \hat{S}_{\Omega_{3}}(r) \hat{b}^{\dag}_{\Omega_{3}I}  (|0 \rangle_{I} \otimes |0 \rangle_{II}) \nonumber \\
    &= \frac{1}{\cosh^{2} r} \sum_{n=0}^{\infty} \sqrt{n+1}\tanh^{n} r|(n+1)_{\Omega} \rangle_I |n_{\Omega} \rangle_{II}. \nonumber
\end{align}

We observe that due to the squeezing behaviour of a non-inertial observer, a single Minkowski mode can be written as a superposition of two Rindler modes. Consequently there exists coherence between the two Rindler modes. Since these two modes are not causally connected, this coherence cannot be experimentally observed. Actually there is no new coherence created in the system.

\begin{figure}
    \includegraphics[width=\columnwidth]{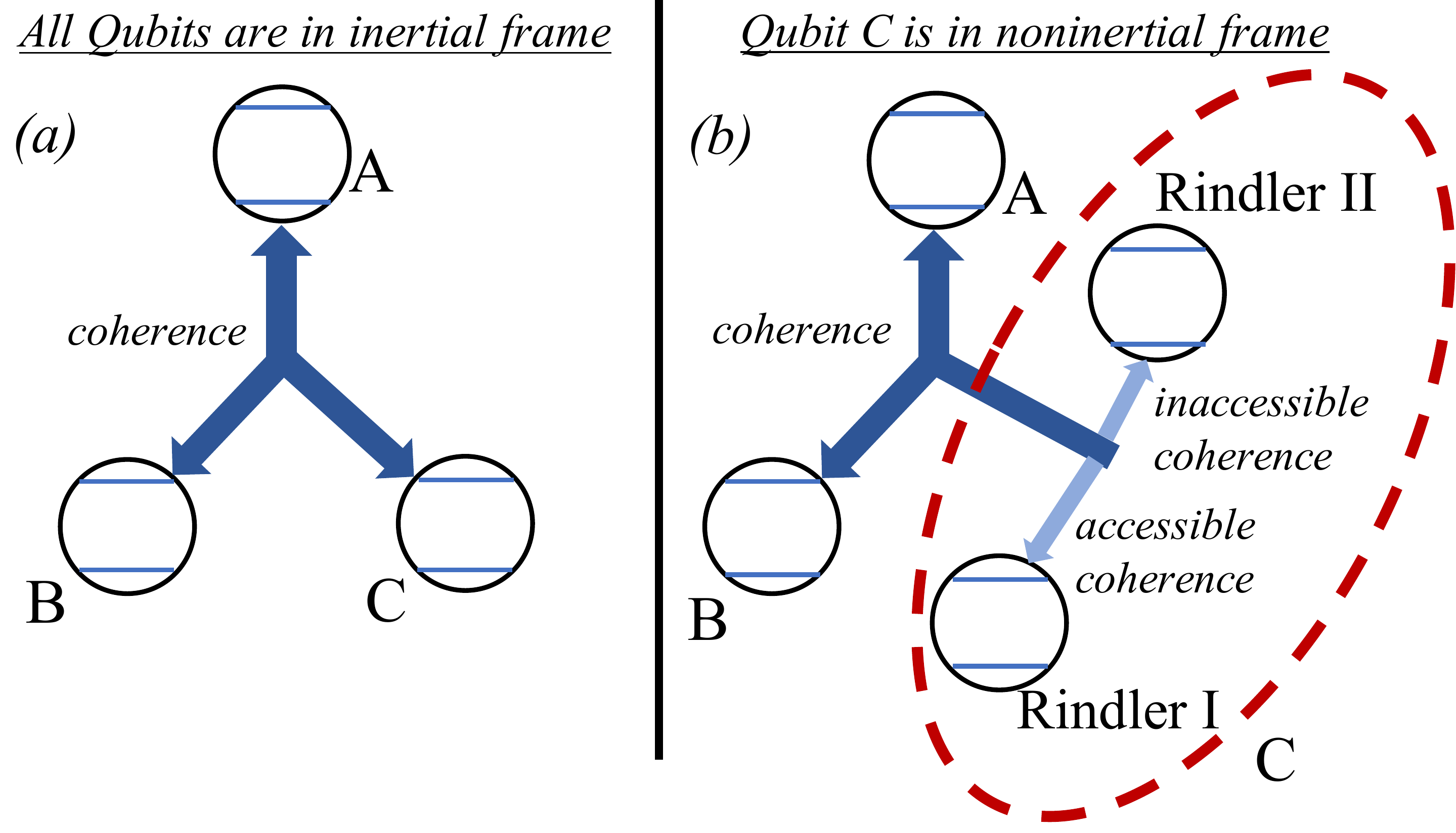} 
    \caption{A schematic diagram of coherence in tripartite is shown where the circles labelled $A$, $B$ and $C$ represent the qubit and the blue arrow between them representing the coherence. Fig (a) contains the coherence in a tripartite system when all the qubits are in inertial frame. The part (b) contains the coherence in the tripartite system when qubit $C$ is under acceleration. Here qubit $C$ enclosed by the Red ellipse with dashed boundary line, is split into two Rindler modes I and II. Consequently the coherence shared by qubit $C$ also is split into two parts. The coherence shared with $C$ of Rindler mode I is the accessible coherence and the coherence shared with $C$ of Rindler mode II is the inaccessible coherence. The lighter shades of blue for the accessible and inaccessible coherence represents their relative strength to the coherence initially present with $C$ and shown in part (a).} \label{fig2}
\end{figure}

To illustrate this let us consider a tripartite system with Alice and Bob at rest, and Charlie moving with constant acceleration. The coherence shared by Charlie with Alice and Bob is split into two parts. One part remains in Rindler mode I and can be experimentally observed, which we call the {\it accessible coherence}. The other part of the coherence which is shared with Rindler mode II cannot be measured, which we refer to as {\it inaccessible coherence}. This inaccessible coherence quantifies the decrease in coherence due to relativistic effects. A schematic diagram explaining the accessible and inaccessible coherence is shown in Fig.~\ref{fig2}.

In a tripartite state when two qubits are accelerated the vacuum ($|000\rangle$) and excited ($|111\rangle$) states can be expressed as,
\begin{widetext}
\begin{align}
    |000 \rangle_{ABC} &= |0 \rangle \otimes |0 \rangle \otimes |0 \rangle \nn \\                    
    &\overset{accl}{\longrightarrow} |0_{\omega_1} \rangle \otimes [\hat{S}_{\Omega_2}(r_{2}) (|0_{\Omega_2} \rangle_{I} \otimes |0_{\Omega_2} \rangle_{II})] \otimes [\hat{S}_{\Omega_3}(r_{1}) (|0_{\Omega_3} \rangle_{I} \otimes |0_{\Omega_3} \rangle_{II})]\nn \\  
    &= |0 \rangle \otimes \frac{1}{\cosh r_{2}} \left[\sum_{m=0}^{\infty} \tanh^{m} r_{2} |m \rangle_{I} |m \rangle_{II}\right] \otimes \left[\frac{1}{\cosh r_{1}} \sum_{n=0}^{\infty} \tanh^{n} r_{1} |n \rangle_{I} |n \rangle_{II}\right], \\
    |111 \rangle_{ABC}  &=  |1 \rangle \otimes |1 \rangle \otimes |1 \rangle \nn  \\
    &\overset{accl}{\longrightarrow} |1_{\omega_1} \rangle \otimes [\hat{S}_{\Omega_2}(r_{2}) \hat{b}^\dagger_{{\Omega_2}{I}} (|0_{\Omega_2} \rangle_{I} \otimes |0_{\Omega_2} \rangle_{II})] \otimes [\hat{S}_{\Omega_3}(r_{1}) \hat{b}^\dagger_{{\Omega_3}{I}}(|0_{\Omega_3} \rangle_{I} \otimes |0_{\Omega_3} \rangle_{II})]\nn \\
    &= |1 \rangle \otimes \left[\frac{1}{\cosh^{2} r_{2}} \sum_{m=0}^{\infty} \sqrt{m+1} \tanh^{m} r_{2} |m+1 \rangle_{I} |m \rangle_{II}\right] \otimes \left[ \frac{1}{\cosh^{2} r_{1}} \sum_{n=0}^{\infty} \sqrt{n+1} \tanh^{n} r_{1} |n+1 \rangle_{I} |n \rangle_{II}\right].
\end{align}
\end{widetext}

The modes $I$ and $II$ correspond to the two causally disconnected regions in Rindler co-ordinates of Minkowski space. Hence mode $II$ is physically inaccessible to Alice, Bob and Charlie and is partially traced out. For an inertial observer, a quantum system in Minkowski space-time is independent of the nature of the observer. But when the observer is moving with constant acceleration, they perceive a thermal bath with temperature proportional to acceleration. This thermal bath is made up of Rindler particles which are associated with the vacuum state of Minkowski space. Due to this thermal bath, the observer perceives decoherence of the quantum system, a phenomenon known as Unruh decoherence.

Quantum coherence is in general measured as the distance between the quantum state under consideration and the closest incoherent state in the same basis. Several measures of quantum coherence \cite{baumgratz2014quantifying,  rana2016trace, shao2015fidelity, radhakrishnan2016distribution} have been presented, but in the first work on quantum coherence by Baumgratz, Cramer and Plenio \cite{baumgratz2014quantifying} two measures were introduced corresponding to the entropic class and geometric class of measures. The relative entropy based measure of quantum coherence belongs to the entropic class, and the $\ell_{1}$ norm of coherence to the geometric class. Here we use the $\ell_{1}$-norm of quantum coherence, defined as,
\begin{align}
    C_{l_1}(\hat\rho,\hat\rho_{d}) = \| \hat\rho- \hat\rho_{d} \|_{l_1}  = \sum_{i \neq j}|\hat\rho_{i,j}|,   
\end{align}
where $\hat{\rho}$ is a given density matrix and $\hat\rho_{d}$ is the associated decohered density matrix defined as,
\begin{align}
    \hat\rho_{d} = \sum_{i} \hat\rho_{i,i} |i \rangle \langle i |.
\end{align}
We can observe that this is equivalent to the sum of the off-diagonal elements of the density matrix. Using interference fringes, the quantum coherence of a physical system can be measured using the robustness of coherence \cite{wang2017directly}, a measure introduced in 
Ref.~\cite{napoli2016robustness}. In Ref.~\cite{piani2016robustness} it was shown that for a pure state the robustness of coherence measure is equal to the $\ell_{1}$-norm of coherence. Hence the $\ell_{1}$-norm of coherence can be directly obtained from the interference fringes, providing a method to compare theoretical results with experimental data.  

\section{Relativistic effects in the SLOCC class of states} \label{sec:rel_eff_slocc}

Based on the local operations and classical communication (LOCC), tripartite entangled quantum states can be divided into two distinct classes, 
namely the GHZ class and the W class. The entanglement distribution is different in these two class of states. In a GHZ state all the entanglement 
vanishes even with the loss of a single qubit whereas in a W-state a finite amount of entanglement is always present even when we loose a single 
qubit. The coherence of these class of states in inertial frames has been discussed in detail in Ref.~\cite{radhakrishnan2019dynamics}. We extend 
upon this by investigating the coherence when some of the qubits of the tripartite system are in a non-inertial reference frame. In particular, 
we probe the loss of coherence due to the acceleration of qubits. 

\subsection{GHZ class}

A GHZ state is maximally entangled and so the loss of just a single qubit results in the complete loss of entanglement and coherence. The general form of a tripartite Greenberger-Horne-Zeilinger (GHZ) state is,
\begin{align}
|\mathrm{GHZ} \rangle_{ABC} = \cos \theta |000 \rangle_{ABC} + \sin \theta |111 \rangle_{ABC},
\end{align}
where $\theta \in [0,2\pi)$ is a parameter generalizing the GHZ state via the bias between the two basis states. On accelerating qubit $C$ its Minkowski modes are replaced with their corresponding Rindler modes and the tripartite states becomes,
\begin{widetext}
\begin{align}
    |\mathrm{GHZ} \rangle_{ABC} &= \frac{1}{\cosh r} \sum_{n=0}^{\infty}  \tanh^{n}r  \bigg[ \cos \theta |00 \rangle  |n \rangle_{I} |n \rangle_{II} + \frac{\sqrt{n+1}}{\cosh r} \; \sin \theta |11 \rangle |n+1 \rangle_{I}  |n \rangle_{II} \bigg]. \label{onequbitacceleratedGHZstate}
\end{align}
\end{widetext}

Here, $|00 \rangle  |n \rangle_{I} |n \rangle_{II}$ and $|11 \rangle |n+1 \rangle_{I}  |n \rangle_{II}$ refer to the quantum state in which the Minkowski modes of the first two qubits are $|ab \rangle$ and the Rindler modes corresponding to the
third qubit is as $|c \rangle_{I} |c \rangle_{II}$. We know that the modes from the Rindler I and II regions are not causally connected, so we can trace out the modes corresponding to Rindler region II and the density matrix of the state with Alice, Bob and Charlie reduces to,
\begin{widetext}
\begin{align}
    \hat\rho_{\mathrm{GHZ}} &= \frac{1}{\cosh^{2} r} \sum_{n=0}^{\infty} \tanh^{2n}r \bigg[ \cos^{2} \theta  |00n \rangle \langle 00n| + \cos \theta \sin \theta \frac{\sqrt{n+1}}{\cosh r} \bigg(|00n \rangle \langle 11 n+1| + |11 n+1 \rangle \langle 00n| \bigg) \nonumber\\
    &+ \frac{n+1}{\cosh^{2} r} \sin^{2} \theta |11 n+1 \rangle \langle 11 n+1| \bigg]. \label{onequbitaccelerateddensitymatrix}
\end{align}
\end{widetext}

The total coherence in the generalised GHZ state is calculated using the $\ell_{1}$-norm measure of coherence. The sum of the off-diagonal elements of Eq.~(\ref{onequbitaccelerateddensitymatrix}) is 
\begin{align}
C(\hat\rho) = \frac{2 \sin \theta \cos \theta}{\cosh^{3} r} \sum_{n=0}^{\infty} \sqrt{n+1} \tanh^{2n} r.
\label{sumoffdiagonalelements}
\end{align}

Using the trigonometric identities,
\begin{align}
    \sum_{n=0}^{\infty} \tanh^{2n} r &=  \cosh^{2} r,\nonumber\\
    \sum_{n=0}^{\infty} (n+1) \tanh^{2n} r &=  \cosh^{4} r,
\end{align}
and the polylogarithm function
\begin{align}
    {\rm Li}_{-1/2}(z) = \sum_{n=0}^{\infty} \sqrt{n+1} (\tanh^{2} r)^{n+1},
    \label{polylogfunction}
\end{align}
where the polylogarithm function is defined as
\begin{align}
 {\rm Li}_{n}(z)  \equiv  \sum_{k=1}^{\infty}  \frac{z^{k}}{k^{n}} = \frac{z}{1^{n}} + \frac{z^{2}}{2^{n}} + \frac{z^{3}}{3^{n}} + \cdots,
\end{align}
the total coherence of a GHZ state when one of the qubits is in a non-inertial frame can be expressed as,
\begin{align}
C(\hat\rho)  = 2 \sin\theta \cos\theta \frac{{\rm Li}_{-1/2}(\tanh^{2} r)}{\sinh^{2} r \, \cosh r}.
\label{onequbitacceleratedcoherence}
\end{align}

When Charlie's qubit is not accelerated (i.e the $r \rightarrow 0$ limit) the corresponding quantum coherence is,
\begin{align}
C (\hat\rho) = 2 \cos \theta \sin \theta.
\end{align}

\begin{figure}
\includegraphics[width=\columnwidth]{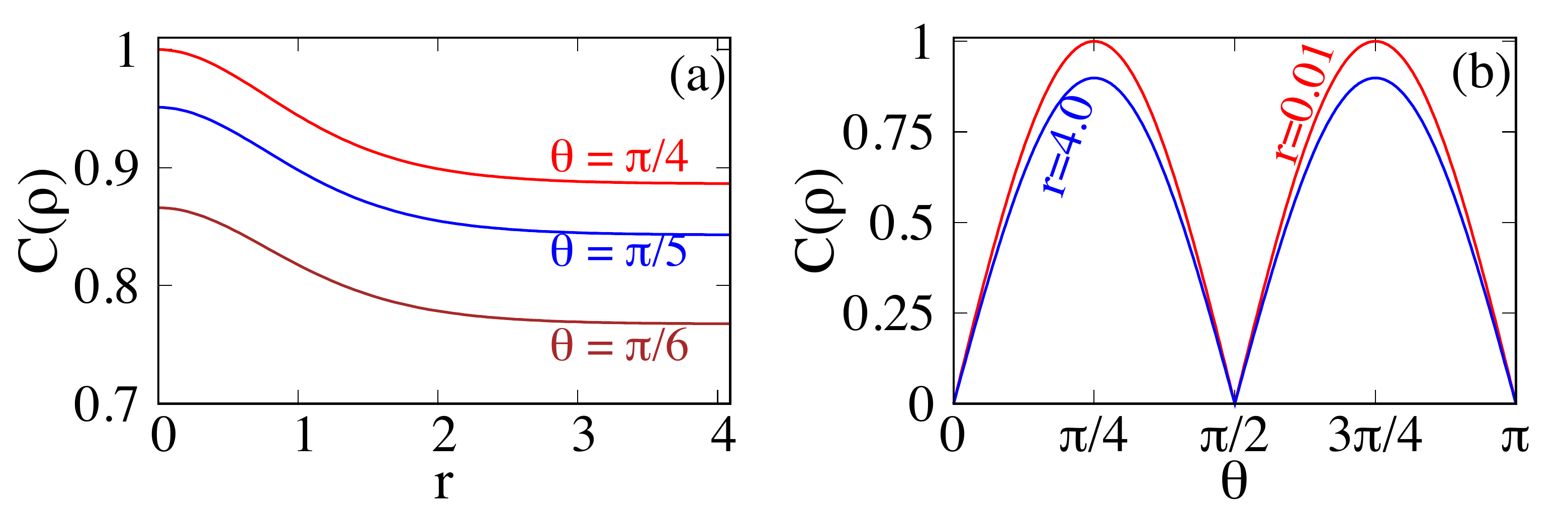}
\caption{The variation of coherence in a non-inertial frame of reference is studied for a GHZ state under the variation of (a) the acceleration parameter $r$ and (b) the generalisation parameter $\theta$.}
\label{fig3}
\end{figure}

The variation of quantum coherence as a function of the acceleration parameter $r$ is shown in Fig.~\ref{fig3}(a) for different $\theta$ values. From the plot for any given value of $\theta$  we notice that quantum coherence has a maximum value when $r=0$, corresponding to the situation where Charlie's qubit has not yet been accelerated. The coherence then decreases with increase in $r$ and saturates to a finite value. The decrease in coherence due to uniform acceleration is because some part of it becomes inaccessible as it lies in a causally disconnected region. The saturation value depends on the value of coherence in the inertial frame ($r=0$). Hence, higher the value of the coherence in the inertial frame, the higher the saturation value of coherence in the non-inertial frame. 

In Fig.~\ref{fig3}(b), we study the change of the quantum coherence as a function of the generalisation parameter $\theta$ for different values of the acceleration parameter $r$. The coherence is maximum at $\theta = (2n+1) \pi/4$ where $n\in\mathbb{Z}$ and is zero at $\theta = n \pi/2$, but the maximum 
value depends on the value of $r$.

To study the situation, when more than one party is under acceleration, we consider the setting where both Bob's and Charlie's qubits are being accelerated. The quantum state when two qubits are accelerated is,
\begin{widetext}
\begin{align}
    |\mathrm{GHZ} \rangle &=  \frac{1}{\cosh r_{1}} \frac{1}{\cosh r_{2}}  \sum_{n,m=0}^{\infty} \tanh^{n}r_{1}   \tanh^{m}r_{2} \bigg[\cos \theta |0\rangle |m\rangle_{I} |m\rangle_{II}  |n\rangle_{I} |n\rangle_{II} \nonumber \\
    &+ \frac{\sqrt{m+1}}{\cosh r_{2}}  \frac{\sqrt{n+1}}{\cosh r_{1}}  \sin \theta \times  |1 \rangle |m+1\rangle_{I} |m \rangle_{II}   
    |n+1\rangle_{I} |n \rangle_{II} \bigg],
\label{twoqubitacceleratedGHZstate}                             
\end{align}
\end{widetext}
where the pairs ($r_{2}$,$m$) and ($r_{1}$,$n$) are the acceleration parameter and mode corresponding to Bob's 
and Charlie's qubits respectively. To construct the density matrix we first trace out the contributions from the causally disconnected Rindler II mode,
\begin{widetext}
    \begin{align}
    \hat\rho  &= \frac{1}{\cosh^{2} r_{1} \cosh^{2} r_{2}}  \sum_{n=0}^{\infty}  \sum_{m=0}^{\infty} \tanh^{2n} r_{1}   \tanh^{2m} r_{2} \bigg[ \cos^{2} \theta |0 \rangle |m\rangle |n\rangle \langle 0| \langle m| \langle n| + \frac{\sqrt{m+1} \sqrt{n+1}}{\cosh r_{1} \cosh r_{2}}  \cos \theta \sin \theta \nonumber \\ 
    & \bigg( |0 \rangle |m \rangle |n \rangle \langle 1 | \langle m+1 | \langle n+1| + |1 \rangle |m+1 \rangle |n+1 \rangle  \langle 0| \langle m | \langle n | \bigg) + \frac{(m+1)(n+1)}{\cosh^{2} r_{1} \cosh^{2} r_{2}} \sin^{2} \theta \nonumber \\
    & |1 \rangle |m+1 \rangle |n+1 \rangle  \langle 1| \langle m+1| \langle n+1| \bigg],  
\label{twoqubitaccelerateddensitymatrix}               
\end{align}
\end{widetext} 

From the density matrix in Eq.~(\ref{twoqubitaccelerateddensitymatrix}) and using the $\ell_{1}$-norm measure we can determine the total quantum coherence of the GHZ state when both Bob's and Charlie's qubits 
are accelerated,
\begin{align}
    C(\hat\rho) = 2 \sin\theta \cos\theta \frac{Li_{-1/2}(\tanh^{2} r_1)}{\sinh^{2} r_1 \cosh r_1} \frac{Li_{-1/2}(\tanh^{2} r_2)}{\sinh^{2} r_{2} \cosh  r_2}, 
\label{twoqubitacceleratedcoherence}
\end{align}
where we find to be a product of two polylogarithm functions one each corresponding to Bob's and Charlie's qubits. If Bob and Charlie have the same acceleration, which would translate to $r_{1} = r_{2} = r$, the coherence would then be,
\begin{align}
    C(\hat{\rho}) = 2 \sin\theta \cos\theta \left[ \frac{Li_{-1/2}(\tanh^{2} r)}{\sinh^{2} r \cosh r}   \right]^{2}.
\end{align}

\begin{figure}
    \includegraphics[width=\columnwidth]{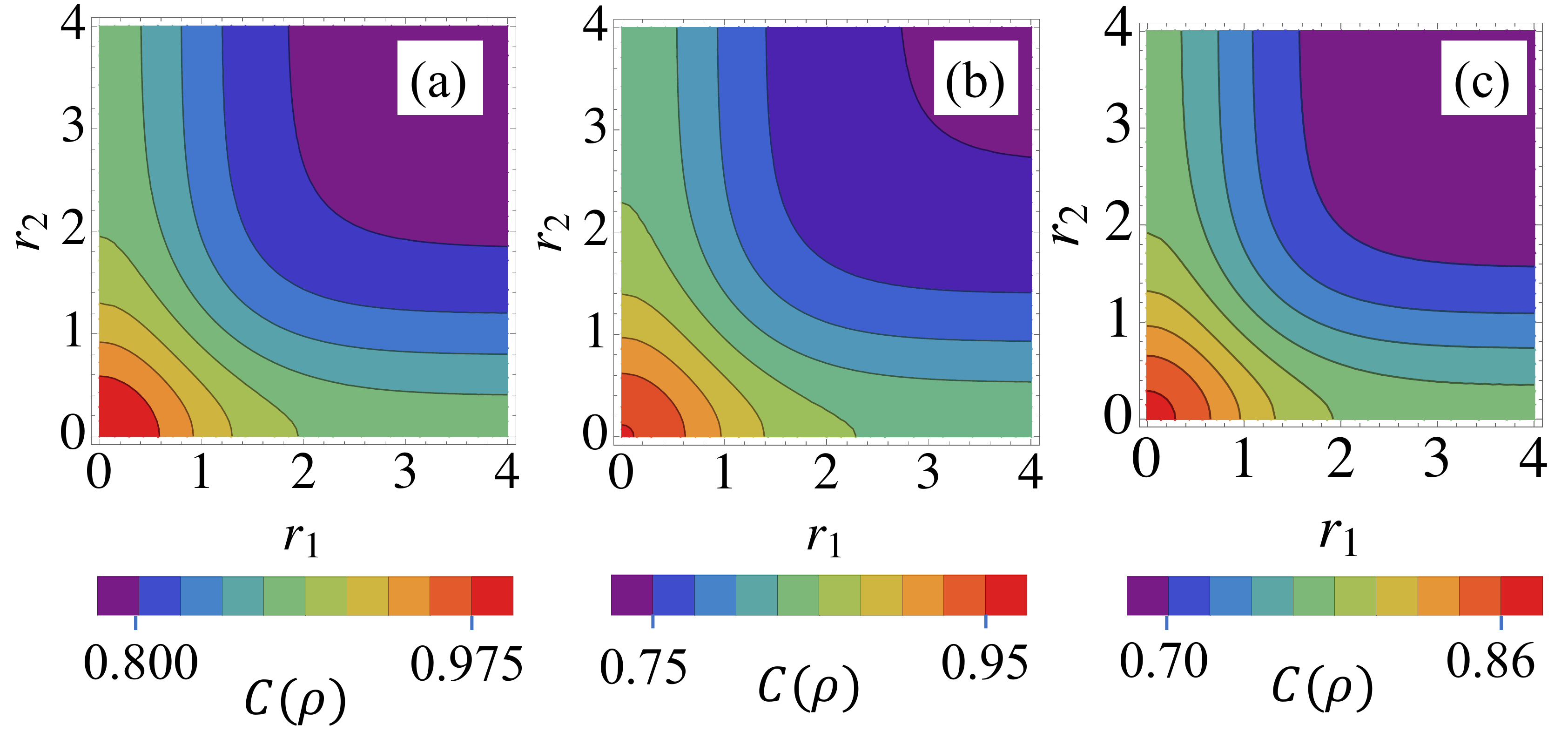} 
    \caption{The variation of quantum coherence of GHZ state with both $r_{1}$ and $r_{2}$ is shown for (a) $\theta = \pi/4$, (b) $\theta =\pi/5$ and (c) $\theta= \pi/6$ respectively.} \label{fig4}
\end{figure}

The variation of quantum coherence with the parameters $r_{1}$ and $r_{2}$ corresponding to Bob's and Charlie's acceleration is shown through the contour plots in Fig.~\ref{fig4}. The contour plots Fig.~\ref{fig4}(a), \ref{fig4}(b) and \ref{fig4}(c) correspond to $\theta$ of $\pi/4$, $\pi/5$ and $\pi/6$ respectively. We find that with the increase in $r_{1}$ and $r_{2}$, the coherence decreases and saturates to a finite value. The maximal value and saturation value of coherence are dependent on the value of $\theta$ the parameter generalizing the GHZ state.  

In literature, the most commonly discussed GHZ state is $(|000 \rangle + |111 \rangle ) / \sqrt{2}$, corresponding to the generalised GHZ state with $\theta = \pi/4$, which has quantum coherence,
\begin{align}
    C(\hat\rho) &= \frac{ Li_{-1/2}(\tanh^{2} r)}{\sinh^{2} r \, \cosh r}~, \nonumber \\
    C(\hat\rho) &= \frac{Li_{-1/2}(\tanh^{2} r_1)}{\sinh^{2} r_1 \cosh r_1} \frac{Li_{-1/2}(\tanh^{2} r_2)}{\sinh^{2} r_{2} \cosh  r_2}.
\end{align}

Finally we note that the two qubit reduced density matrices corresponding to the GHZ state are diagonal in nature and hence are incoherent. Thus all the coherences in the GHZ state are genuinely multipartite in nature and the loss of even one qubit removes all the coherence in the system.

In Ref.~\cite{fuentes2005alice}, the authors considered a two qubit system of which one qubit is moving with a constant acceleration.  The entanglement of the bipartite system vanished in the infinite acceleration limit. It is well known that entanglement is just one of the different types of quantum correlations. Out of the different quantifiers of quantum correlations, quantum discord measures the total quantum correlations in the system. A study of quantum discord in relativistic system \cite{datta2009quantum} show that total quantum correlations does not vanish in the infinite acceleration limit. Hence a bipartite system in the infinite acceleration limit has quantum correlations beyond the entanglement type. An investigation on accelerating tripartite systems
was carried out in Ref.~\cite{hwang2011tripartite}, where the GHZ and W-states were characterised using the $\pi$-tangle measure of entanglement. The entanglement of the tripartite system did not go to zero in the infinite acceleration limit, rather it showed a decrease initially and then attained a saturation value.  
This is in stark contrast to the behaviour of the bipartite entanglement. In Ref.~\cite{hwang2011tripartite} it was suggested that the incomplete definition of $\pi$-tangle could be the reason as to why the tripartite entanglement shows a different qualitative behaviour when compared to the bipartite entanglement as measured in Ref.~\cite{fuentes2005alice}. Through the present work we observe that the quantum coherence of the GHZ state is qualitatively similar to the change in its tripartite entanglement. But here the saturation is due to the non-vanishing behaviour of local quantum correlations. However for quantum coherence, the $\ell_{1}$-norm measures the entire quantum coherence and the saturation value is a physical feature of the system. In the case of entanglement, the saturation value may not be an actual physical feature since $\pi$-tangle is inadequate to measure entanglement. 

Next we can consider the three qubit GHZ state given in Eq.~(\ref{GHZdefinition}) in which two qubits are under uniform acceleration. The GHZ state can be rewritten as $(|0 \mathbf{0} \rangle + |1 \mathbf{1} \rangle)/\sqrt{2}$, 
where $|\mathbf{0} \rangle = |00 \rangle$ and $|\mathbf{1} \rangle = |11 \rangle$ are redundantly-encoded logical qubits. Since the GHZ state can be written in terms of the Bell state, we would expect the coherences of both these states to be equal. But the coherence of the GHZ state when the logical qubit is accelerated is,
\begin{align}
    C_{\ell_{1}} =  [ Li_{-1/2} (\tanh^{2} r) / (\sinh^{2} r  \cosh r) ]^{2},
\end{align}
which implies that $C(|\psi_{Bell} \rangle )  \neq  C(|\psi_{GHZ} \rangle)$ an relativistic system. This is because 
in the GHZ state, the logical non-inertial qubit is bigger than the regular qubit considered in the Bell state. So, we find that the quantum coherence coupled with uniform acceleration helps in distinguishing between Bell states and GHZ states.  

\begin{figure}
    \includegraphics[width=\columnwidth]{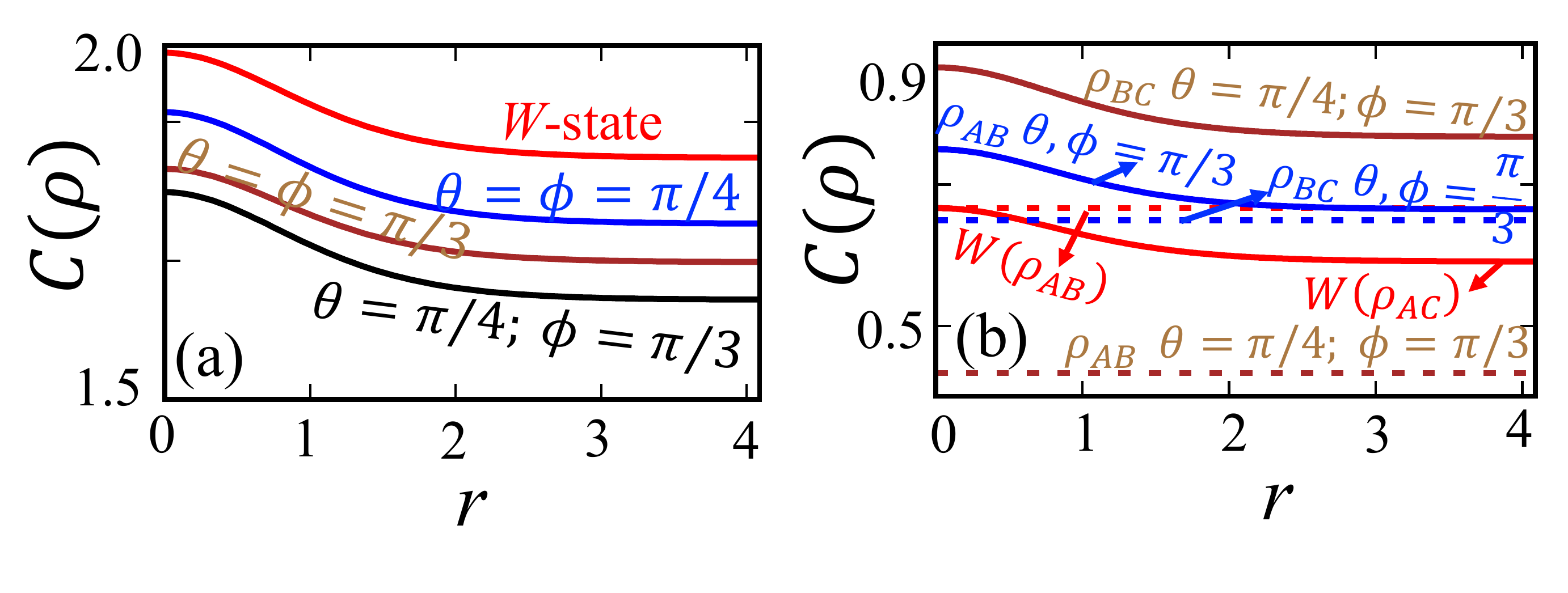}
    \caption{The variation of quantum coherence as a function of $r$ for fixed values of $\theta$ and $\phi$ is given for (a) the tripartite generalized W-state and (b) the two qubit reduced systems $\rho_{AB}$, $\rho_{BC}$ and $\rho_{AC}$.  Here the notation $W(\rho_{AB})$ means the reduced state $\rho_{AB}$ corresponding to the symmetric $W$-state.}
\label{fig5}
\end{figure}

\subsection{W class}

The W-states exhibit non-maximal multipartite entanglement which is not locally equivalent to GHZ-type entanglement, and may therefore be considered a distinct entanglement class. Unlike GHZ states, W-states with large numbers of qubits are highly robust against qubit loss. The generalised form of the tripartite W-state is,
\begin{align}
    |W \rangle_g = \sin\theta \cos\phi|100\rangle + \sin\theta \sin\phi|010\rangle + \cos\theta|001\rangle, \label{generalizedWstate}
\end{align} 
where $\theta = [0,\pi]$ and $\phi = [0, 2 \pi)$ are the parameters generalising the W-state. 

First let us consider the situation when only one qubit is in a non-inertial frame of reference. Towards this end, we consider the situation when Charlie's qubit is in a non-inertial frame while Alice and Bob's qubits are in an inertial frame. Naturally the Minkowski modes of Charlie's qubit are replaced by their corresponding Rindler modes. It is well known that the Minkowski region can be divided into two causally disconnected Rindler modes. So, we trace out the second Rindler region from the quantum state, since it is inaccessible for measuerment. The total quantum coherence is then measured using the  $\ell_{1}$-norm of coherence,
\begin{align}
    C(\hat\rho) &= 2 \sin\theta \cos\theta (\sin\phi+\cos\phi) \frac{ Li_{-1/2}(\tanh^{2} r)}{\sinh^{2} r \, \cosh r} \nonumber \\
    & +2 \sin^2\theta \sin\phi \cos\phi~.  \label{Wstatesinglequbitacceleration}    
\end{align}

\begin{figure}
    \includegraphics[width=\columnwidth]{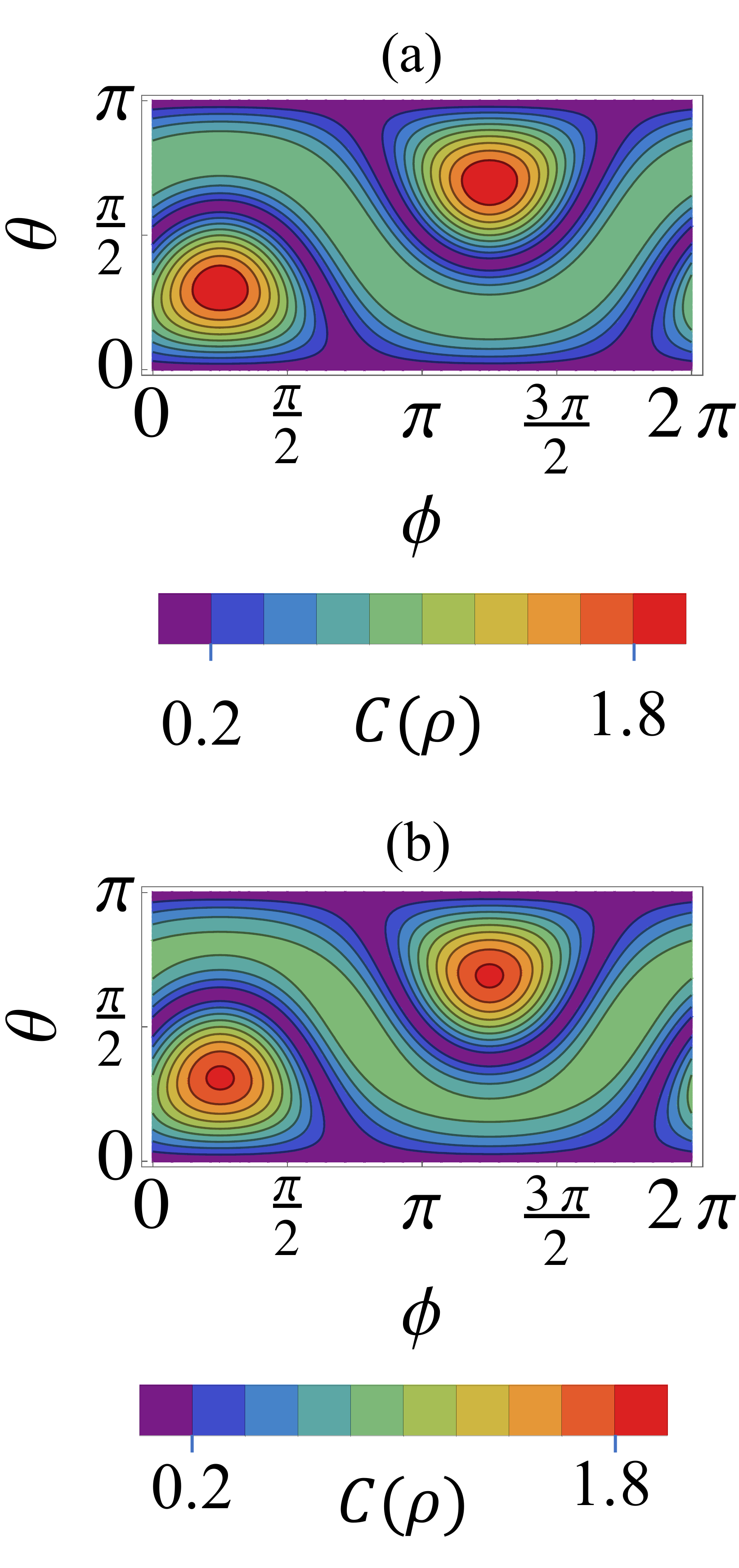} 
    \caption{Contour plot of the Quantum coherence $C(\rho)$ as a function of $\theta$ and $\phi$ of the generalised W-state when one qubit is in a non-inertial frame is shown for (a) $r=0.01$ and (b) $r=4.0$.} \label{fig6}
\end{figure}

To understand the relativistic effects on a quantum state with bipartite distribution, we compute the reduced density matrices corresponding to the $W$-state. Here, we can trace out either Alice or Bob's qubit and the quantum coherence corresponding to the reduced state is,
\begin{align}
    C(\hat\rho_{BC}) = C(\hat\rho_{AC}) = 2 \sin\theta \cos\theta \sin\phi \frac{Li_{-1/2}(\tanh^2 r)}{\sinh^2 r \cosh r}.
\label{2qubitreducedWstateacceleration1qubit}
\end{align}

The reduced density matrix obtained after tracing out Charlie's qubit does not have any effects of non-inertial nature and the coherence of the joint state of Alice and Bob is,
\begin{align}
    C(\hat\rho_{AB}) = 2 \sin^2\theta \sin\phi \cos\phi,
\end{align}
which is the expected standard result. The most common form the three qubit W-state is the tripartite state of the form
$|W \rangle = \frac{1}{\sqrt{3}} (|001 \rangle + |010 \rangle + |100 \rangle )$, and of all the generalised W-states it has the maximum amount of coherence and entanglement. Also the total coherence and entanglement are distributed in a symmetric manner. The quantum coherence of a tripartite system when Charlie's qubit is accelerated is ,
\begin{align}
C(\hat\rho)  = \frac{4 \, Li_{-1/2}(\tanh^{2} r)}{3 \cosh r \,  \sinh^{2} r} + \frac{2}{3}.
\end{align}

In the $r \rightarrow 0$ limit (inertial limit), it reduces to the standard value of $C(\hat\rho) = 2$. The change in coherence of the tripartite generalised W-state is given as a function of the acceleration parameter $r$ for different values of $\theta$ and $\phi$ in Fig.~\ref{fig5}(a). From the plots we notice that
the quantum coherence is initially maximum at $r=0$ and then it decreases and reaches a saturation value at large values of $r$.  The decrease in coherence is due to the bifurcation of the Minkowski mode into two Rindler modes, of which one is disconnected from the rest of the system. Any coherence due to this mode is inaccessible to measurements. The saturation value is the amount of accessible coherence which can never be lost to relativistic motion. The maximal value and saturation value are dependent on the generalisation parameter. The $|W \rangle = \frac{1}{\sqrt{3}} (|001 \rangle + |010 \rangle + |100 \rangle )$ state has the maximal coherence at $r=0$ and also has the maximal saturation value. The quantum coherence of the reduced density matrices of the generalised W-state is shown in Fig.~\ref{fig5}(b). The quantum coherence of the reduced state $\hat\rho_{AB}$ is a constant since the accelerated 
qubit $C$ has already been traced out. In the case of $\hat\rho_{BC}$ we find that the coherence is maximal at $r=0$ and saturates to a finite value at large accelerations. The curve represents the coherence of $\hat\rho_{BC}$ and the dashed lines denote the coherence in the state $\hat\rho_{AB}$. In Fig.~\ref{fig6} the contour plot shows the variation of quantum coherence as a function of the generalisation parameter $\theta$ and $\phi$ for the acceleration parameters $r=0.01$ and $r=4.0$.    

Next we look at the case where two qubits, the ones corresponding to Bob and Charlie are being accelerated. Adopting a similar procedure of replacing the Minkowski states by Rindler modes and tracing out the Rindler second mode, we measure the total quantum coherence of the system. The quantum coherence of the tripartite system is,
\begin{align}
    C(\hat\rho) &= 2 \sin\theta\cos\theta\sin\phi   \nonumber \\  
    & \times \frac{Li_{-1/2}(\tanh^2 r_1)}{\sinh^2 r_1 \cosh r_1}    \frac{Li_{-1/2}(\tanh^2 r_2)}{\sinh^2 r_2 \cosh r_2} \nonumber \\  
    & +2 \sin\theta \cos\theta \cos\phi \frac{Li_{-1/2}(\tanh^{2} r_1)}{\sinh^{2} r_1 \cosh r_1} \nonumber \\
    & +2 \sin^2\theta \sin\phi \cos\phi \frac{Li_{-1/2}(\tanh^{2} r_2)}{\sinh^{2} r_{2} \cosh  r_2}. \label{Wstatetwoqubitacceleration}    
\end{align}

\begin{figure}
\includegraphics[width=\columnwidth]{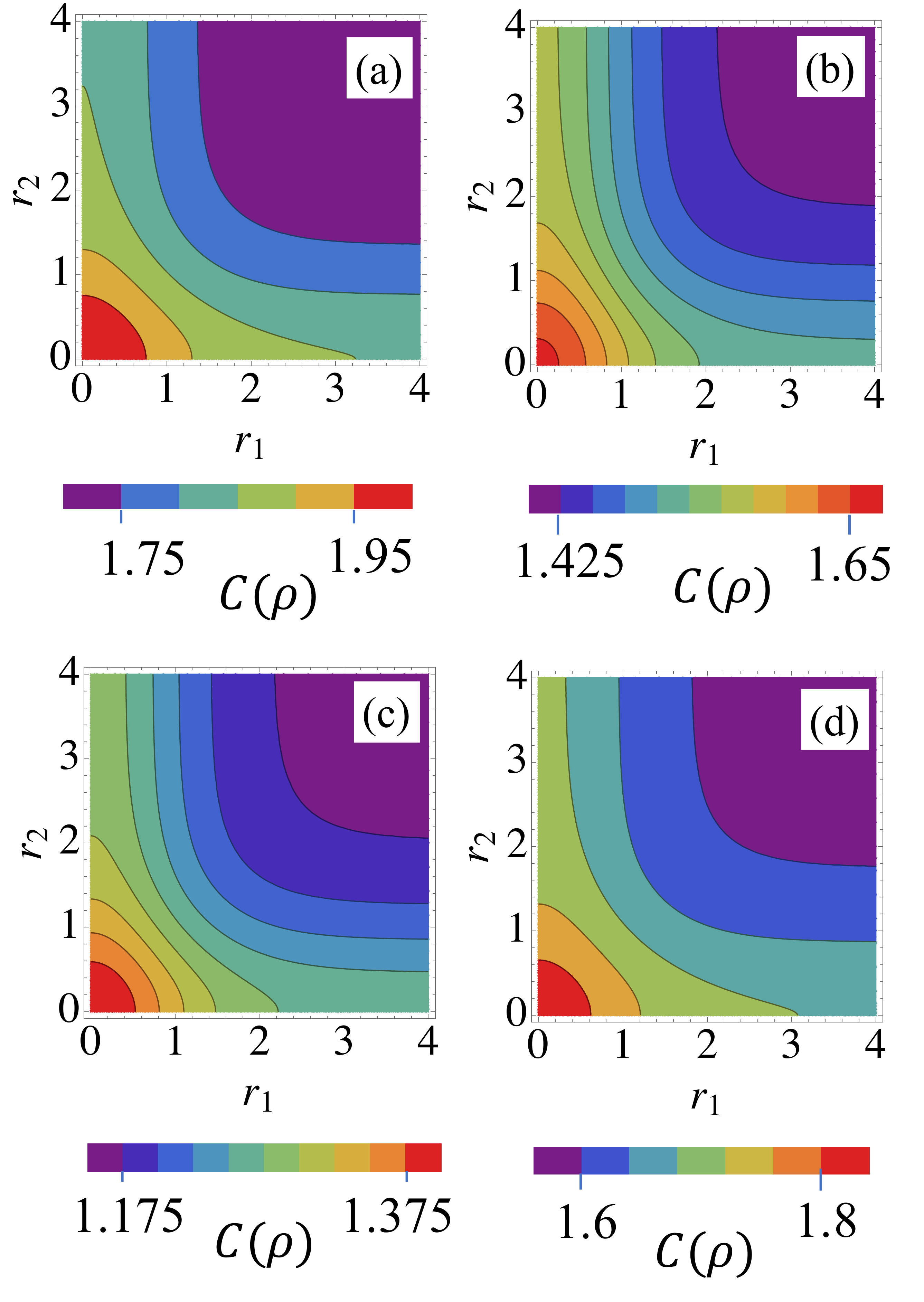}
\caption{In the generalised W-state when both Bob's and Charlie's qubits are accelerated, the variation of quantum coherence as a function of the acceleration parameters $r_{1}$ and $r_{2}$ is given for (a) $W$-state, (b) $\theta = \pi/5$ , $\phi =\pi/5$, (c) $\theta = \pi/6$ , $\phi =\pi/6$ and (d) $\theta = \pi/3$ , $\phi =\pi/6$.}
\label{fig7}
\end{figure}

From Eq.~(\ref{Wstatetwoqubitacceleration}) we notice that the quantum coherence is a sum of three terms. The first contains the non-inertial decoherence from both Bob's and Charlie's qubit. The non-inertial contributions from Bob and Charlie appear in the second and third terms respectively. At the bipartite level, there are two possibilities namely: (i) when Alice's qubit is traced out; (ii) when either Bob's qubit or Charlie's qubit is traced out. The quantum coherence corresponding to these different situations 
are,
\begin{align}
    C(\hat\rho_{BC}) &= 2 \sin\theta \cos\theta \sin\phi \frac{Li_{-1/2}(\tanh^2 r_1)}{\sinh^2 r_1 \cosh r_1} \nonumber \\
    & \times \frac{Li_{-1/2}(\tanh^2 r_2)}{\sinh^2 r_2 \cosh r_2}, 
\label{BobCharlie2qubitcoherence}   \\    
    C(\hat\rho_{AC}) &= 2 \sin\theta \cos\theta \sin\phi \frac{Li_{-1/2}(\tanh^2 r_1)}{\sinh^2 r_1 \cosh r_1},
\label{AliceCharlie2qubitcoherence}  \\                 
    C(\hat\rho_{AB}) &= 2 \sin^2\theta \sin\phi \cos\phi \frac{Li_{-1/2}(\tanh^2 r_2)}{\sinh^2 r_2 \cosh r_2}.
\label{AliceBob2qubitcoherence}  
\end{align}

Here Eq.~(\ref{BobCharlie2qubitcoherence}) refers to the situation where Alice's qubit is traced out and in the resulting bipartite system both the qubits are in non-inertial frames. When Bob's qubit is traced out the total quantum coherence is given through Eq.~(\ref{AliceCharlie2qubitcoherence}), and similarly when 
Charlie's qubit is traced out Eq.~(\ref{AliceBob2qubitcoherence})  gives the quantum coherence of the resulting bipartite state. For the $W$-state of the form$ \frac{1}{\sqrt{3}} (|001 \rangle + |010 \rangle + |100 \rangle )$ when both Bob's and Charlie's qubits are 
accelerated the total quantum coherence of the system is,
\begin{align}
    C(\hat\rho) &= \frac{2}{3} \frac{Li_{-1/2}(\tanh^{2} r_{1})}{\sinh^{2} r_{1} \cosh r_{1}} \frac{Li_{-1/2}(\tanh^{2} r_{2})}{\sinh^{2} r_{2} \cosh  r_{2}} \nonumber\\
    & + \frac{2}{3} \frac{Li_{-1/2}(\tanh^{2} r_{1})}{\sinh^{2} r_{1} \cosh r_{1}} + \frac{2}{3} \frac{Li_{-1/2}(\tanh^{2} r_{2})}{\sinh^{2} r_{2} \cosh r_{2}}. \label{twoqubitaccelerationwstate}
\end{align}

Eq.~(\ref{twoqubitaccelerationwstate}) reduces to the standard value of $C(\hat\rho) = 2$ for $r_{1},r_{2} \rightarrow 0$ and the single qubit acceleration limit when either $r_{1}$ or $r_{2}$ tends to zero. In the contour plot shown in Fig.~\ref{fig7} we analyse the variation of the quantum coherence as a function of $r_{1}$ and $r_{2}$ namely Charlie's and Bob's acceleration parameter for different values of the generalisation parameters $\theta$ and $\phi$. We find that coherence is maximum when $r_{1}$ and $r_{2}$ are zero and decreases with increase in their value.  

A very interesting limiting case of the generalised W-state occurs when $\theta = \pi/2$ and $\phi = \pi/4$. The generalised W-state corresponding to this value is $(|100 \rangle + |010 \rangle)/\sqrt{2}$. We can observe that this quantum state is of the biseparable form AB-C, where the AB pair is entangled and C is separable. In this state Charlie's qubit is accelerated, the total quantum coherence is $C = 1$. This is because the only coherence contribution in the system comes because of the correlation 
between Alice and Bob's qubit. On the contrary, when Bob is in a non-inertial frame of reference, the total quantum coherence in the system is,
\begin{align}
    C(\hat\rho) = \frac{Li_{-1/2}(\tanh^{2} r)}{\sinh^{2} r \cosh r}. \label{biseparablestate1and2qubitacceleration} 
\end{align}
When both Bob and Charlie are in non-inertial frames, the resulting coherence is same as when Bob is in a non-inertial frame. This is because Charlie's qubit does not contribute to the coherence in the system. 

For the W-state, the tripartite entanglement measured by the $\pi$-tangle saturates at a finite value \cite{hwang2011tripartite}. This is in stark contrast with the results obtained for the bipartite entanglement in Ref.~\cite{fuentes2005alice}. This is again due to the insufficiency in using $\pi$-tangle as a measure of entanglement. For a $W$-state, the variation of the quantum coherence due to uniform acceleration is similar to the change in the entanglement of the system. But the $\ell_{1}$-norm measure of coherence is a complete measure unlike the $\pi$-tangle measure of entanglement used in Ref.~\cite{hwang2011tripartite}. The  saturation nature of quantum coherence is due to the presence of non-classical correlations at relativistic velocities. This result holds for single qubit as well as two qubit accelerated systems. 

\section{Separable states} \label{sec:separable}

The SLOCC class of classification applies to entangled states. But in general quantum states may not be entangled. An example of a separable state is $|000 \rangle$ and for this state $\hat\rho = \hat\rho_{d}$ in the computational basis and so it is an incoherent state. Here we would like to mention that one of the crucial properties of quantum coherence is that it is a basis-dependent
quantity. Hence studying a separable quantum state $|+++ \rangle$ where $|+ \rangle  = (|0\rangle + |1 \rangle)/\sqrt{2}$ in the $\sigma^{z}$-basis we find quantum coherence in the system. To find the relativistic effects, we accelerate either one or two of these qubits. For the accelerated qubits we replace the Minkowski states by their corresponding Rindler modes. We already know that there are two causally disconnected Rindler modes, so we trace out one of the modes and compute the quantum coherence of the rest of the system. Here when only Charlie's qubit is accelerated the quantum coherence
computed using the $\ell_{1}$-norm measure is,
\begin{align}
    C(\hat\rho) = 3 + \frac{4 \; Li_{-1/2}(\tanh^{2} r)}{\cosh r  \sinh^{2} r}. \label{separablestateonequbitaccelerated}
\end{align}
In the $r \rightarrow 0$ limit, the inertial value of the $\ell_{1}$-norm coherence of the system is recovered. When both Charlie's and Bob's qubit are accelerated the quantum coherence is,
\begin{align}
    C(\hat\rho) &= 1 + \frac{2 \; Li_{-1/2}(\tanh^{2} r_{1})}{\cosh r_{1}  \sinh^{2} r_{1}} + \frac{2 \; Li_{-1/2}(\tanh^{2} r_{2})}{\cosh r_{2}  \sinh^{2} r_{2}}  \nonumber \\
    & + \frac{2 \; Li_{-1/2}(\tanh^{2} r_{1})}{\cosh r_{1}  \sinh^{2} r_{1}} \; \frac{Li_{-1/2}(\tanh^{2} r_{2})}{\cosh r_{2}  \sinh^{2} r_{2}}. \label{separablestatetwoqubitaccelerated}
\end{align}

From this result, in the appropriate limiting conditions we can recover the single qubit result as well as the inertial value.  

The coherence measured in the GHZ state, W-state and the separable $|+++ \rangle$ state is the total amount of quantum coherence present in the states. But the type of coherence in the GHZ and W-states is fundamentally different from the type of coherence present in the $|+++ \rangle$ state. In the GHZ and W-states, the coherence arises due to the correlation between the qubits. Meanwhile the coherence in the $|+++ \rangle$ state is because of the superposition between the levels within each qubit. These two fundamentally different forms of coherences were identified in Ref.~\cite{radhakrishnan2016distribution}. The coherence arising due to the correlation between the qubits is the global coherence of the system and the coherence resulting from the superposition of the levels within a qubit are called local coherence. These two forms of coherences are complementary to each other in a sense that the increase in the global coherence causes a decrease in the local coherence and vice versa. With increase in the acceleration the quantum coherence decreases and it saturates at a finite value. To understand this for the $|+++ \rangle$ state, we can consider the single qubit system $|+ \rangle$ system which we can write as,
\begin{align} 
    |+ \rangle = \sum_{n=0}^{\infty}  \frac{\tanh^{n} r }{\sqrt{2}  \cosh r} \left( |n \rangle_{I}  |n \rangle_{II} + \frac{\sqrt{n+1}}{\cosh r} |n+1 \rangle_{I} |n \rangle_{II} \right).
\end{align}

Here we can observe that the initial superposition between two levels is spread out between the four modes. Of these two modes corresponding to Rindler region II are traced out and hence there is a loss of superposition and consequently a loss of coherence. This loss is the inaccessible coherence when the system undergoes uniform acceleration.  

The GHZ and W-states represents one extreme where the global 
coherence is maximum with zero local coherence and the \mbox{$|+++\rangle$} denotes the other extreme where the local coherence is maximum with no global coherence. But there are some pure quantum states in which both these types of coherence coexist. For such states we can quantify the global and local coherence using the formula,
\begin{align}
    C_{G} &= \| \hat\rho - \pi(\hat\rho) \|_{l_1},\label{globalcoherence} \\
    C_{L} &= \|\pi(\hat\rho) - [\pi(\hat\rho)]_{d} \|_{l_1}. \label{localcoherence}
\end{align}

Here $\pi(\hat\rho) = \hat\rho_{1}\otimes\dots\otimes \hat\rho_{N}$ is the product state of the density matrix $\hat\rho$ where the reduced density matrix $\hat\rho_{1} = {\rm Tr}_{(2,...,N)} \ \hat\rho$. The quantum state  $[\pi(\hat\rho)]_{d}$ is the decohered density matrix corresponding to $\pi(\hat\rho)$ the product state. The total quantum coherence of the system $C_{T}$ measures coherence contributions from correlations as well as local superpositions. In Ref.~\cite{radhakrishnan2016distribution} a trade-off was observed 
between the global coherence and local coherence. 

\begin{figure*}
    \includegraphics[width=2.0\columnwidth]{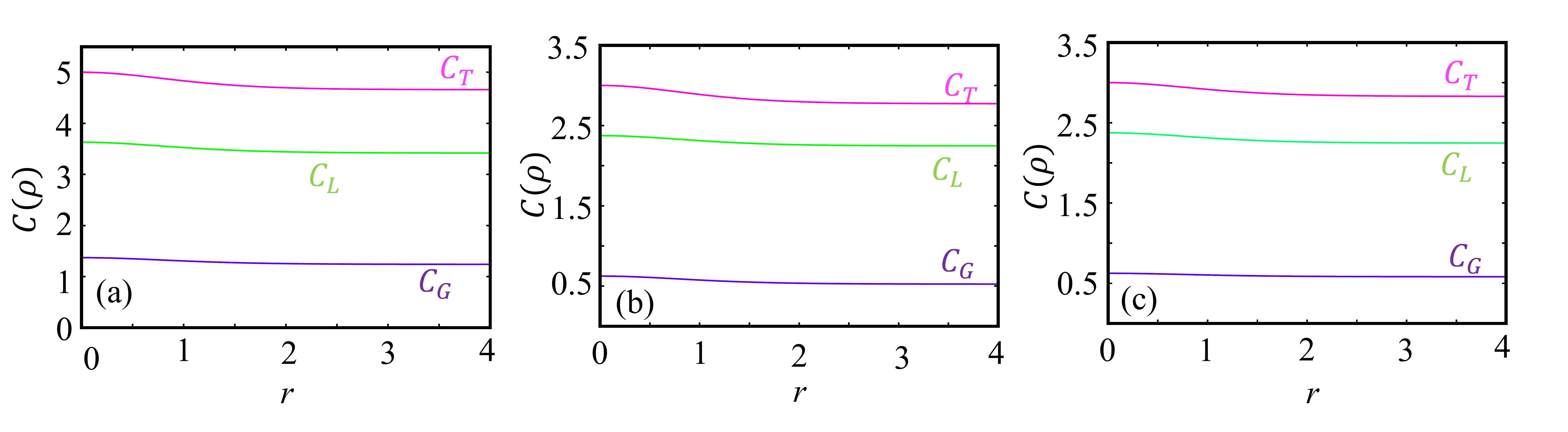} 
    \caption{A plot of the total coherence $(C_{T})$, global coherence $(C_{G})$ and local coherence $(C_{L})$ for (a) the $W \bar{W}$ state, (b) Star state when the central qubit is accelerating and (c) Star state when the peripheral qubit is accelerating.} \label{fig8}
\end{figure*}

\section{Tripartite pure states with local and global coherence} \label{sec:tripartite}

In this section we study the non-inertial effects when both local and global coherences are present in the system. Towards this end we analyze the quantum coherence of tripartite $W \bar{W}$ and star states. The $W \bar{W}$ state is a symmetric tripartite state with 
both local and global coherence. On the contrary the star state is an asymmetric state with both local and global coherences.  

\subsection{$W \bar{W}$ state}

The three qubit $W \bar{W}$ state \cite{AKRajagopal_WWbarstate2012} is a linear superposition of the tripartite $W$ state and the $\bar{W}$ state as shown below,
\begin{align}
    |W \bar{W} \rangle &= \frac{1}{\sqrt{2}} (|W \rangle  + | \bar{W} \rangle ) \label{wwbarstate},\nonumber \\
    |W \rangle &= \frac{1}{\sqrt{3}} ( |001 \rangle + |010 \rangle + |100 \rangle ),\nonumber \\
    |\bar{W} \rangle &= \frac{1}{\sqrt{3}} ( |011 \rangle + |101 \rangle + |110 \rangle ). 
\end{align}

First we consider the situation, where a single qubit is in a noninertial frame that is to say only Charlie's qubit is being accelerated. After tracing out one of the Rindler region, the total quantum coherence of the resulting density matrix is,
\begin{align}
    C_{T}(\hat\rho) = 2 + \frac{3 \; Li_{-1/2}(\tanh^{2} r) }{\cosh  r \sinh^{2} r}.\label{wwbar1qubitacceleration}
\end{align}
In the $r \rightarrow 0$ limit the quantum coherence recovers the inertial limit. 

The single qubit reduced density matrices of the $W \bar{W}$ state in Eq.~(\ref{wwbarstate}) are,
\begin{align}
    \hat\rho_{A}  =  \hat\rho_{B}  = \hat\rho_{C} = 
    \begin{pmatrix}
    \frac{1}{2}  &  \frac{1}{3} \\
    \frac{1}{3}  &  \frac{1}{2}
    \end{pmatrix}.
\end{align}
From the reduced density matrix we notice that the single qubit state has a finite amount of quantum coherence. Consequently the product state $\hat\rho_{A} \otimes \hat\rho_{B} \otimes \hat\rho_{C}$ will also have some coherence which is the local coherence of the system. Using Equations (\ref{globalcoherence}) and (\ref{localcoherence}) we find the global and local coherence when one of the qubit is in non-inertial frame,
\begin{align}
    C_{G} (\hat\rho) &=  \frac{2}{9} + \frac{31 \; Li_{-1/2}(\tanh^{2} r)}{27 \cosh r  \sinh^{2} r} \label{localcoherencewwbar1qubitacceleration}, \\
    C_{L} (\hat\rho) &=  \frac{16}{9} + \frac{50 \; Li_{-1/2}(\tanh^{2} r)}{27 \cosh r  \sinh^{2} r}. \label{globalcoherencewwbar1qubitacceleration} 
\end{align}

The behaviour of the total coherence, global coherence and local coherence are shown in Fig.~\ref{fig8}(a). From the plots we find that all the different forms of coherences show identical behaviour and they decrease with increase in acceleration and attain a saturation value at higher values of acceleration. The results show that both global and local coherence have similar decoherence properties. Hence Unruh decoherence cannot distinguish between the two forms of coherence {\it viz} the one arising due to the correlations between the qubits and the quantumness due to the superposition between the levels within a qubit.  

To study the relativistic effects when two qubits are in noninertial frames, we uniformly accelerate both Bob's and Charlie's qubits. The total quantum coherence of the $W \bar{W}$ state in this scenario is,
\begin{align}
    C_{T} (\hat\rho) &= \frac{4}{6} + \frac{8 \; Li_{-1/2}(\tanh^{2} r_{1}) }{6 \cosh  r_{1}  \sinh^{2} r_{1}} + \frac{8 \; Li_{-1/2}(\tanh^{2} r_{2}) }{6 \cosh  r_{2}  \sinh^{2} r_{2}} \nonumber \\
    &+ \frac{10  Li_{-1/2}(\tanh^{2} r_{1}) \;  Li_{-1/2}(\tanh^{2} r_{2}) }{6 \cosh  r_{1}  \sinh^{2} r_{1} \;  \cosh  r_{2}  \sinh^{2} r_{2}}. \label{totalcoherencewwbar2qubitaccelerated}
\end{align}

The total coherence is present as both global coherence arising from inter-qubit correlations and also as local coherence coming from intraqubit superpositions. The measured values of the global and local coherence of the system are,
\begin{align}
    C_{G} (\hat\rho)  &= \frac{2 \; Li_{-1/2}(\tanh^{2} r_1)}{ 9 \cosh r_1 \sinh^{2} r_1} + \frac{2\; Li_{-1/2}(\tanh^{2} r_2)}{9 \cosh r_2 \sinh^{2} r_2} \nonumber \\ 
    &+ \frac{25 \; Li_{-1/2}(\tanh^{2} r_1) \; Li_{-1/2}(\tanh^{2} r_2)}{ 27 \cosh r_1  \sinh^{2} r_1 \; \cosh r_2 \sinh^{2} r_2}, 
\label{globalcoherencewwbar2qubitaccelerated}    \\
    C_{L} (\hat\rho) &= \frac{2}{3} +  \frac{10 \; Li_{-1/2}(\tanh^{2} r_1)}{ 9 \cosh r_1 \sinh^{2} r_1} + \frac{10\; Li_{-1/2}(\tanh^{2} r_2)}{9 \cosh r_2 \sinh^{2} r_2} \nonumber \\ 
    & + \frac{20 \; Li_{-1/2}(\tanh^{2} r_1) \; Li_{-1/2}(\tanh^{2} r_2)}{ 27 \cosh r_1 \sinh^{2} r_1 \; \cosh r_2 \sinh^{2} r_2}. 
\label{localcoherencewwbar2qubitaccelerated}
\end{align}

We find that the total coherence, global coherence and local coherence reduce to their respective inertial values when $r_{1}, r_{2} \rightarrow 0$.  Also we find that the total coherence is a sum of the global and local coherence for the $\ell_{1}$-norm of coherence. 

\subsection{Star state}

A star state is an asymmetric quantum state \cite{Plesch_starstate2003, Plesch_starstateII_2003}, so called because a central qubit is entangled with the peripheral qubits collectively, but upon tracing out the central qubit the peripheral ones have a separable form,
\begin{align}
    \hat\rho_{B,C}=\mathrm{tr}_A(\ket{S}_{A,B,C}\bra{S}_{A,B,C}) = \hat\rho_B\otimes\hat\rho_C.
\end{align}
In the tripartite star state we have a central qubit $A$ which is entangled to other qubits $B$ and $C$ individually. The qubits $B$ and $C$ are not entangled with each other and are referred to as peripheral qubits. The form of the tripartite star state is,
\begin{align}
    |S \rangle  = \frac{1}{2} (|000 \rangle + |100 \rangle + |101 \rangle + |111 \rangle). \label{starstateexpression}
\end{align}
When the central qubit $A$ is accelerated, the total quantum coherence in the system is,
\begin{align}
    C_{T} (\hat\rho) = 1 + \frac{2 \; Li_{-1/2}(\tanh^{2} r)}{\cosh r  \sinh^{2} r}. \label{sstotalcoherence1qubitcentral}
\end{align}
The star state has both global coherence and local coherence and they are 
\begin{align}
    C_{G} (\hat\rho) &= \frac{-1}{4} + \frac{7 \; Li_{-1/2}(\tanh^{2} r)}{ 8 \cosh r  \sinh^{2} r},\label{ssglobalcoherence1qubitcentral} \\
    C_{L} (\hat\rho) &= \frac{5}{4} + \frac{9 \; Li_{-1/2}(\tanh^{2} r)}{ 8 \cosh r  \sinh^{2} r}. \label{sslocalcoherence1qubitcentral}
\end{align}

When the peripheral qubit $B$ is being accelerated, the total coherence of the system calculated using the $\ell_{1}$-norm of coherence is,
\begin{align}
    C_{T} (\hat\rho) = \frac{3}{2} + \frac{3 \; Li_{-1/2}(\tanh^{2} r)}{2 \cosh r \sinh^{2} r}.\label{sstotalcoherence1qubitperipheral}
\end{align}
The corresponding global of the system is,
\begin{align}
    C_{G} (\hat\rho) = \frac{1}{4} + \frac{3 \; Li_{-1/2}(\tanh^{2} r)}{ 8 \cosh r  \sinh^{2} r}. \label{ssglobalcoherence1qubitperipheral}
\end{align}

The local coherence of the system on accelerating the peripheral qubit is same as the local coherence when the central qubit is being under acceleration, the expression for which is given in Eq.~(\ref{sslocalcoherence1qubitcentral}). In both the cases namely when the central qubit and the peripheral qubit in the $r \rightarrow 0$ limit the results corresponding to the inertial frame are recovered. For the star states, the variation of quantum coherence is qualitatively similar when both the central and peripheral qubits are accelerated. The global local and total coherence have a maximum value in the inertial frame and decreases with acceleration and attain a saturation value for large values of the acceleration as shown in Fig.~\ref{fig8}. From the analysis of the two different situations where either the central qubit or the peripheral qubit is being accelerated, we find that both global and local coherence have similar decoherence properties. This shows that Unruh decoherence affects all kinds of quantumness equally.  

Next we look at the situation where two qubits are in non-inertial frames of reference. Due to the asymmetry of the star states, there are two possibilities for this situation {\it viz}: (i) when a peripheral qubit and a central qubit are accelerated ; (ii) when both the peripheral qubits are accelerated. The first situation can be analysed when both Bob's and Charlie's qubits are simultaneously accelerated. The total coherence of the system in this case is,
\begin{align}
    C_{T} (\hat\rho) &=  \frac{1}{2} + \frac{Li_{-1/2}(\tanh^{2} r_{1})}{\cosh r_{1}  \sinh^{2} r_{1}} + \frac{1}{2} \; \frac{Li_{-1/2}(\tanh^{2} r_{2})}{\cosh r_{2}  \sinh^{2} r_{2}} \nonumber \\
    &+ \frac{Li_{-1/2}(\tanh^{2} r_{1})}{\cosh r_{1}  \sinh^{2} r_{1}} \; \frac{Li_{-1/2}(\tanh^{2} r_{2})}{\cosh r_{2}  \sinh^{2} r_{2}}. \label{totalcoherence2qubitaccelerationpc}
\end{align}

Under the same conditions the global and local coherence of the star state are,
\begin{align}
    C_{G}(\hat\rho) &=  \frac{1}{2} +  \frac{1 \; Li_{-1/2}(\tanh^{2} r_1)}{ 4 \cosh r_1 \sinh^{2} r_1} - \frac{1 \; Li_{-1/2}(\tanh^{2} r_1)}{ 4 \cosh r_2 \sinh^{2} r_2} \nonumber \\
    & + \frac{5 \; Li_{-1/2}(\tanh^{2} r_1) \; Li_{-1/2}(\tanh^{2} r_2)}{ 8 \cosh r_1  \sinh^{2} r_1 \; \cosh r_2 \sinh^{2} r_2}, \label{globalcoherence2qubitaccelerationpc} \\
    C_{L} (\hat\rho) &= \frac{1}{2} + \frac{3 \; Li_{-1/2}(\tanh^{2} r_1)}{ 4 \cosh r_1 \sinh^{2} r_1} + \frac{3\; Li_{-1/2}(\tanh^{2} r_2)}{4 \cosh r_2 \sinh^{2} r_2} \nonumber \\ 
    &+ \frac{3 \; Li_{-1/2}(\tanh^{2} r_1) \; Li_{-1/2}(\tanh^{2} r_2)}{ 8 \cosh r_1  \sinh^{2} r_1 \; \cosh r_2 \sinh^{2} r_2}. \label{localcoherence2qubitaccelerationpc}     
\end{align}

In the second case when Alice and Bob's qubits are being accelerated, the total coherence of the system is,
\begin{align}
    C_{T}(\hat\rho) &= \frac{1}{2} + \frac{Li_{-1/2}(\tanh^{2} r_{1})}{\cosh r_{1}  \sinh^{2} r_{1}} + \frac{Li_{-1/2}(\tanh^{2} r_{2})}{\cosh r_{2}  \sinh^{2} r_{2}} \nonumber \\
    & + \frac{1}{2} \; \frac{Li_{-1/2}(\tanh^{2} r_{1})}{\cosh r_{1}  \sinh^{2} r_{1}} \; \frac{Li_{-1/2}(\tanh^{2} r_{2})}{\cosh r_{2}  \sinh^{2} r_{2}}. \label{totalcoherence2qubitaccelerationpp}
\end{align}

The corresponding global and local coherences are,
\begin{align}
    C_{G} (\hat\rho) &= \frac{ \; Li_{-1/2}(\tanh^{2} r_1)}{ 4 \cosh r_1 \sinh^{2} r_1} + \frac{ Li_{-1/2}(\tanh^{2} r_2)}{4 \cosh r_2 \sinh^{2} r_2} \nonumber \\ 
    & + \frac{ Li_{-1/2}(\tanh^{2} r_1) \; Li_{-1/2}(\tanh^{2} r_2)}{ 8 \cosh r_1  \sinh^{2} r_1 \; \cosh r_2 \sinh^{2} r_2},  \label{globalcoherence2qubitaccelerationpp} \\
    C_{L}(\hat\rho) &=  \frac{1}{2} + \frac{3 \;  Li_{-1/2}(\tanh^{2} r_1)}{4 \cosh r_1 \sinh^{2} r_1} +  \frac{3 \;  Li_{-1/2}(\tanh^{2} r_2)}{4 \cosh r_2 \sinh^{2} r_2} \nonumber \\
    & + \frac{3 \; Li_{-1/2}(\tanh^{2} r_1) \; Li_{-1/2}(\tanh^{2} r_2)}{8 \cosh r_1  \sinh^{2} r_1 \; \cosh r_2 \sinh^{2} r_2}.
\end{align}

All the different quantum coherences attain their inertial values in the limit $r \rightarrow 0$.  From the expressions of the total, global and local coherence we find that the $C_{T} = C_{G} + C_{L}$ for all the different cases of star states.

In the tripartite systems like $W \bar{W}$-states and star states, the total coherence in the system is distributed as global and local coherence. Here by the word global coherence we mean coherence arising due to all the types of non-classical correlations (both local and entanglement type) between the qubits. Hence the global coherence does not fall to zero in the infinite acceleration limit and rather saturates to a finite value  due to the presence of local correlations. The effect of non-inertial motion on local coherence has not been investigated before.  From our results we can see that the intra-qubit superpositions do not completely vanish in the infinite acceleration limit which is a very interesting result. In the tripartite systems when either one or two of the parties are in non-inertial frames of reference, the accelerated qubits get split into two modes corresponding to the two Rindler regions.  The degradation of quantum coherence is because part of the coherence becomes inaccessible to experimental measurement. Both the global and local coherence decay at the same rate. Hence we find that the acceleration affects the interqubit and intraqubit properties in the same manner.  

\begin{figure}
    \includegraphics[width=\columnwidth]{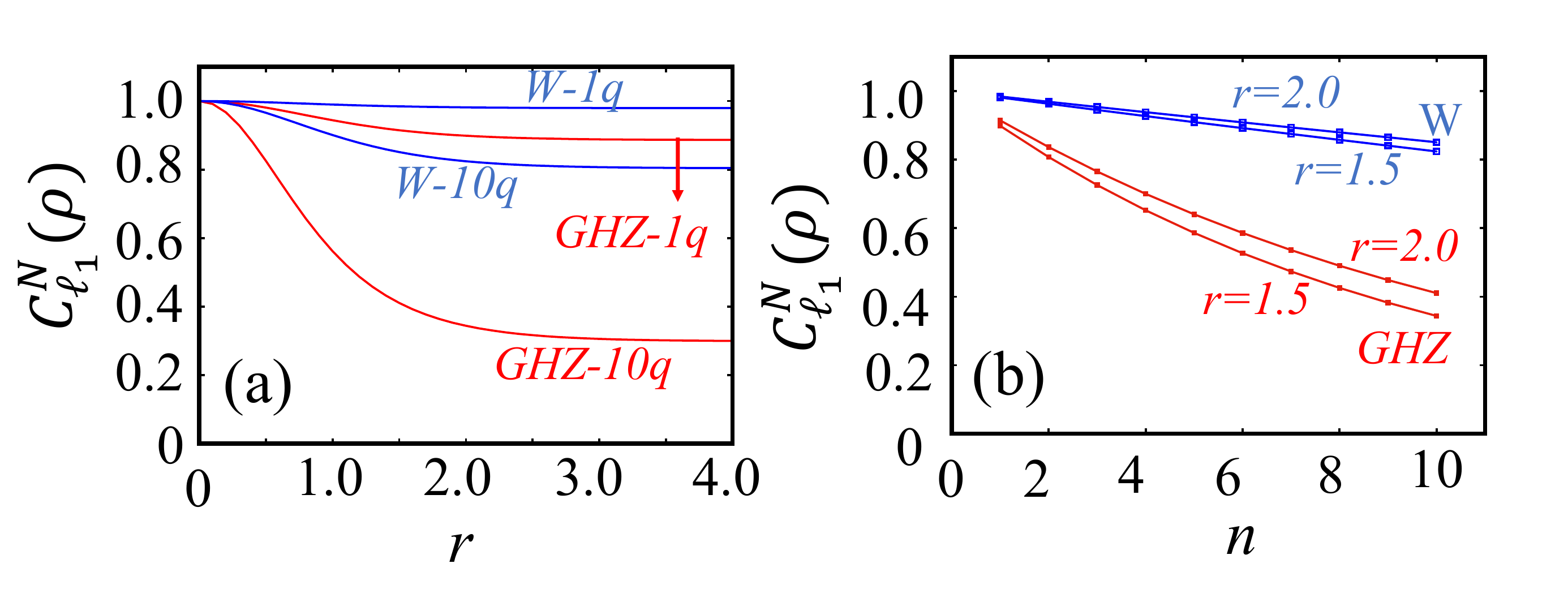} 
    \caption{A plot of the variation of the normalized decoherence (a) with the acceleration parameter $r$ and (b) with the number of accelerated qubits $n$ for a fixed values of $r=1.5$ and $r=2.0$. The total number of qubits in the system is $N=11$.} \label{fig9}
\end{figure}

\section{Applications} \label{sec:apps}

\subsection{Relativistic effects on multipartite state distributed over a network}

Let us consider a network of satellites sharing a multipartite quantum state. An important study, could be about the effect of acceleration on the quantum coherence of the multipartite system. To understand this, we measure the quantum coherence of the $N$-partite GHZ and W-state. In the 
$N$ partite state, we have $n$ number of uniformly accelerated qubits. First let us consider a $N$-partite GHZ state, 
\begin{align}
    |\mathrm{GHZ} \rangle = \frac{1}{\sqrt{2}}(|0 \rangle^{\otimes N} + |1 \rangle^{\otimes N}).
\end{align}
The total quantum coherence is $C_{\ell_{1}} (\hat\rho) = 1$ for this state. Out of the $N$-qubits if $n$ qubits undergo acceleration, the quantum coherence of the state is,
\begin{align}
    C(\hat\rho) = \prod_{k=1}^{n} \frac{Li_{-1/2}(\tanh^{2} r_{k})}{\sinh^{2} r_{k}  \cosh r_{k}}.
\end{align}

Let $C_{k} (\hat\rho)$ be the quantum coherence of the $N$-partite system when only the $k^{th}$ qubit is being accelerated. Using this the coherence of the $N$-partite system with $n$ accelerating qubits can be written as,
\begin{align}
C(\hat\rho) =  \prod_{k=1}^{n} C_{k} (\hat\rho).
\end{align}

When all the satellites have the same acceleration, the quantum coherence simplifies to $C(\hat\rho) = [ C_{k} (\hat\rho) ]^{n}$. From this result we can see that the quantum coherence of the $N$-partite GHZ state falls exponentially with the increase in the number of accelerating satellites. This exponential fall is a manifestation of genuinely multipartite form of quantum coherence where the loss of even a single qubit makes the state completely incoherent. Hence a small acceleration of each of the $m$ qubits leads to a huge loss of coherence in the whole system. To put this 
into a numerical perspective let us consider a $11$ qubit GHZ state without any of the qubits under acceleration. The quantum coherence of such a system is $C_{\ell_{1}} (\hat\rho) = 1$ and let us consider $r =2.0$ for each of the qubit. The coherence when only one qubit is accelerating is $C_{\ell_{1}} (\hat\rho) = 0.8988$. Now when $10$ qubits start accelerating, we can see that the total coherence reduced to $C_{\ell_{1}} (\hat\rho) = (0.8988)^{10} = 0.3439$. For very large values of $n$, the quantum coherence $C_{\ell_{1}} \rightarrow 0$. Since the reduction of quantum coherence depends on its distribution in multipartite system, the results obtained here might hold for entanglement as well. This is because coherence and entanglement are distributed in a similar manner. So we can expect entanglement to fall exponentially similar to quantum coherence.  Hence when quantum information is shared between a network of satellites, the quantum coherence is a function of the number of accelerating satellites and the amount of acceleration.  

Next we consider the $N$-partite W-state,
\begin{align}
    | W \rangle = \frac{1}{\sqrt{N}}  (|00 \cdots 0 1 \rangle + |00 \cdots 10 \rangle + \cdots + |10 \cdots 00 \rangle).
\end{align}
In the non-relativistic scenario the total quantum coherence of this state in the non-inertial frame of reference is $C_{\ell_{1}} (\hat\rho) = N-1$. 
Here if $n$ qubits start accelerating, the quantum coherence of the $N$-partite state changes to,
\begin{align}
    C_{\ell_{1}} (\hat\rho) &= \frac{2}{N}   \sum\limits_{\substack{1 \leq i,j \leq n  \\ i <  j }} \frac{Li_{-1/2}(\tanh^{2} r_{i})}{\sinh^{2} r_{i}  \cosh r_{i}} \frac{Li_{-1/2}(\tanh^{2} r_{j})}{\sinh^{2} r_{j}  \cosh r_{j}} \nonumber \\
    &+ \frac{2}{N} (N-n) \sum_{i=1}^{N}  \frac{Li_{-1/2}(\tanh^{2} r_{i})}{\sinh^{2} r_{i}  \cosh r_{i}}  \nonumber \\ 
    &+  \frac{(N-n)(N-(n+1))}{N}.\label{wstatenaccel}  
\end{align}

In the large $N$ limit, when $n \approx N$ the first term which is a product of two polylog functions dominates over the second and the third term. Hence we would see the quantum coherence falling as polynomial function of second order. When $n \ll N$ in the large $N$ limit, the third term dominates over the first and second term. Consequently the 
decrease of quantum coherence due to the relativistic effects will be minimal under such situations. If acceleration is the same for all the satellites then we have,
\begin{align}
    C_{\ell_{1}} (\hat\rho) &= \frac{2}{N} \frac{n(n-1)}{2}   \left[   \frac{Li_{-1/2}(\tanh^{2} r_{i})}{\sinh^{2} r_{i}  \cosh r_{i}}  \right]^{2}  \nonumber \\
    &+ \frac{2}{N} (N-n) n \left[   \frac{Li_{-1/2}(\tanh^{2} r_{i})} {\sinh^{2} r_{i}  \cosh r_{i}}  \right]  \nonumber \\
    &+ \frac{ (N-n) (N-(n+1)) } {N}.
\end{align}

Since the quantum coherence of the $N$-partite W-state is $(N-1)$, it is not possible to compare it with the coherence loss of the GHZ state. To make a comparison between the coherence loss of the GHZ and W-states we define,
\begin{align}
    C_{\ell_{1}}^{N} (\hat\rho) = \frac{C_{\ell_{1}}^{R} (\hat\rho)}{C_{\ell_{1}}^{NR} (\hat\rho)},
\end{align}
where $C_{\ell_{1}}^{N}$ is the normalized amount of quantum coherence and $C_{\ell_{1}}^{R}$ is the amount of relativistic coherence and $C_{\ell_{1}}^{NR}$ is the quantum coherence of the non-relativistic state. For the GHZ state $C_{\ell_{1}}^{NR}=1$ and so $C_{\ell_{1}}^{N} (\hat\rho) = C_{\ell_{1}}^{R} (\hat\rho)$. In the case of the W-state, for $r=2.0$, $C_{\ell_{1}}^{R} (\hat\rho) = 8.2430$ and $C_{\ell_{1}}^{NR}=N-1$. The normalized amount of quantum coherence in the system when $10$ out of $11$ qubits are being accelerated is $C_{\ell_{1}}^{N}=0.8243$. Hence, for the same number of quantum states and acceleration, we find that the quantum coherence of W-state is more robust to Unruh decoherence when compared with GHZ state. This implies that in a multipartite system sharing quantum coherence in a bipartite manner protects it from decoherence due to relativistic effects.  

Figs.~\ref{fig9}(a) and \ref{fig9}(b) depicts the relativistic quantum coherence of the system. In Fig.~\ref{fig9}(a) we plot the change in quantum coherence with $r$ for both GHZ and W-states.  Here we consider two situations where one qubit is being accelerated or $10$ qubits are being accelerated. In the first case, where one qubit is under acceleration the loss of coherence is minimal. Compared to the first case, we find that the coherence loss is much higher when $10$ qubits are accelerated. Also we find that among these states, the W-state with increasing accelerating qubits is more robust to decoherence. In Fig.~\ref{fig9} (b) we plot the coherence as a function of the number of accelerating qubits. We find that the loss of coherence is higher with increase in the number of accelerating qubits. Also for a given number of accelerating qubits, the GHZ states experience a higher loss of coherence.

\subsection{Coherence degradation at infinite acceleration}
The entanglement degradation in the highly relativistic situation has been explained in Ref. ~\cite{fuentes2005alice} 
in the context of a bipartite systems.  In Ref.~\cite{fuentes2005alice}, the authors consider the two qubit case where Alice is an 
inertial observer and Bob being non-inertial with uniform acceleration.  In the infinite acceleration limit, non-inertial Bob is 
close to the Rindler horizon Ref.~\cite{fuentes2005alice}, which can be considered as the event horizon from the perspective of a 
black hole. 
It is well known that a spherical non-rotating static black hole can be described using a Schwarzchild space-time.  This can be approximated by Rindler co-ordinate in Minkowski space-time in the infinite acceleration limit \cite{Wald}.  
Hence in the existing scenario, we 
consider Alice and Bob to be close to the event horizon, such that inertial Alice falls into a static black hole and accelerating
Bob close to event horizon escapes from falling into it.

Similarly our work investigates the quantum coherence degradation in tripartite systems due to relativistic effects.  
Here we consider three qubits, one each in the possession of Alice, Bob and Charlie.  Degradation of quantum coherence has been 
observed when either a single qubit or two qubits are being accelerated.  From our results in Sec. ~\ref{sec:rel_eff_slocc}, 
for both GHZ and W-states we notice a decrease in quantum coherence, but the coherence freezes at a finite amount in the 
infinite acceleration limit.  The entire coherence in the GHZ and W-state is due to the correlations between the qubits. 
In a similar context to understand the relativistic effects on local coherence we study the $|+++ \rangle$ separable state.
The qualitative behavior of the local coherence in the $|+++ \rangle$ state is similar to that of the global coherence of the GHZ
and W-states.  Hence we conclude that when an inertial observer falls into a black hole, the quantum correlations and 
quantum superposition shared with the non-inertial observer does not vanish completely.

\section{Results \& discussion} \label{sec:discussion}

The relativistic effects on the quantum coherence of a multipartite system is investigated using the $\ell_{1}$-norm measure of coherence. 
Initially all the qubits are in an inertial frame and can be described using Minkowski space-time coordinates.  
When some of the qubits are under acceleration, the Minkowski space corresponding to them is divided into two 
causally disconnected regions. From the Minkowski modes, the Rindler modes can be obtained via the Unruh modes.  When we consider
a Gaussian Minkowski smearing function it can be approximated by a single Unruh mode. Under the monochromatic approximation we can describe the states using Rindler co-ordinates in the Minkowski space time. Hence the quantum coherence 
initially present in the Minkowski co-ordinates is distributed between the two Rindler co-ordinates of which only one is 
experimentally accessible. The coherence which can be measured in the accessible Rindler region is called the accessible 
coherence. The coherence corresponding to the other Rindler region is called the inaccessible coherence since it cannot 
be experimentally measured.

First we investigate the tripartite quantum systems. Based on the SLOCC classification, the tripartite states are classified 
into the GHZ class and the W classes. For both the GHZ and W-states we find that the quantum coherence decreases with increase 
in acceleration and attains a saturation value for very high accelerations. Here we note that coherence can exist due to 
correlations between qubits which is known as global coherence. Also coherence may arise due to superposition between the 
levels within a qubit, referred to as local coherence. The global coherence is the only coherence present in the GHZ and 
the W-states.

Next we consider a separable state of the form $|+++ \rangle$, to understand the relativistic effects on local coherence. 
The local coherence also decreases with increase in acceleration and saturates at higher acceleration. The local coherence 
and the global coherence are complementary to each other and the total coherence is a combination of these two types of 
coherences. Some tripartite quantum states have both local and global coherence. As an example we consider two such quantum 
states namely $W \bar{W}$ and star states. In the $W \bar{W}$ state, the global coherence is distributed equally between 
the three qubits. There is an asymmetric distribution of quantum coherence in the star states. For both these states, 
all the different forms of coherence namely the global, local and total coherence decreases with an increase in acceleration.  

\begin{figure}
    \includegraphics[width=\columnwidth]{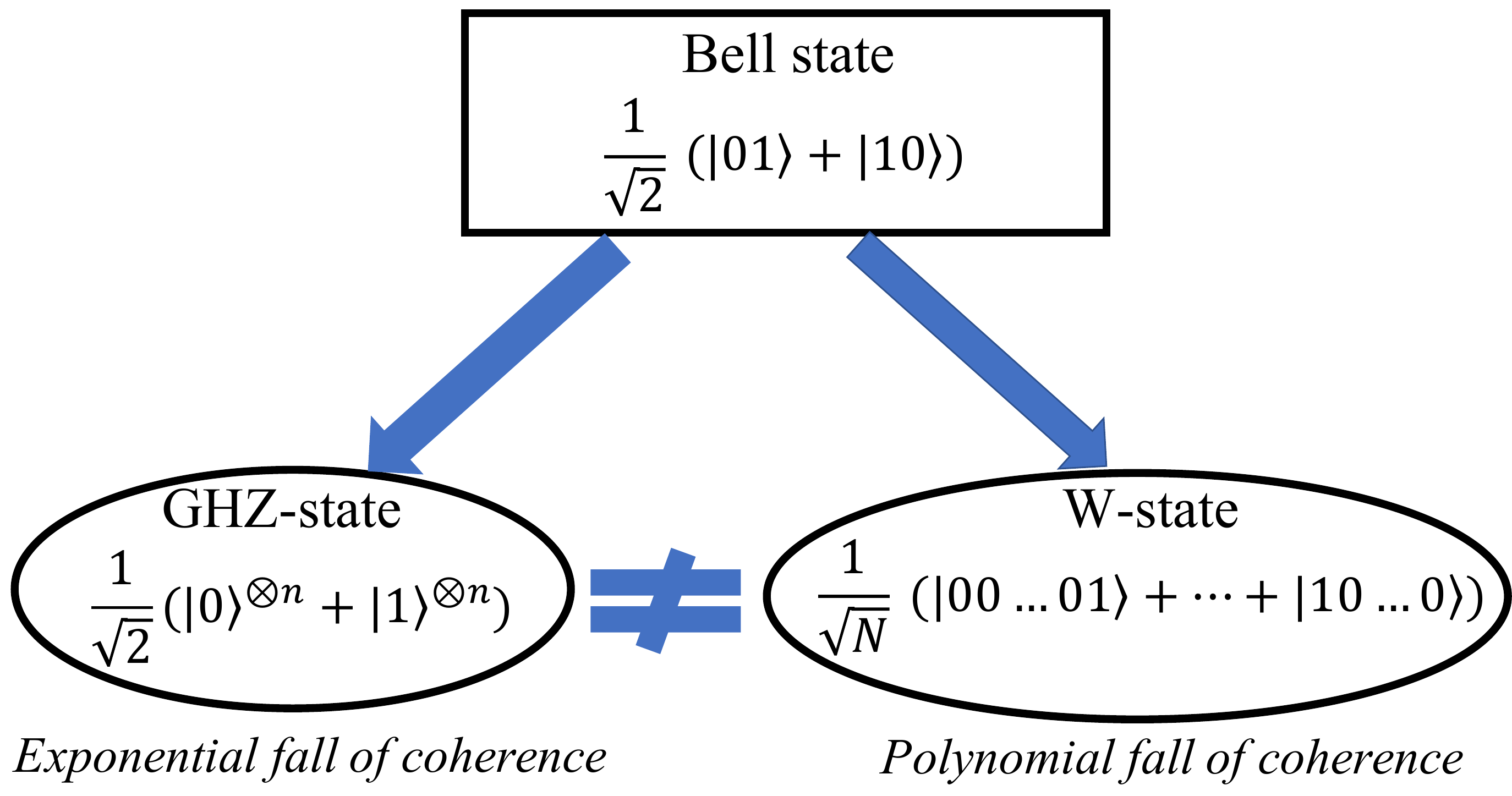}
    \caption{A schematic illustration of the bifurcation of the Bell state which displays linear loss of coherence 
    into two extreme classes of states {\it viz.} GHZ class with exponential loss of coherence and W class with polynomial 
    loss of coherence.} \label{fig10}
\end{figure}

The entanglement of the bipartite system decreases due to relativistic effects and reaches zero in the infinite 
acceleration limit \cite{fuentes2005alice}. But the tripartite entanglement measured in Ref.~\cite{hwang2011tripartite} 
using the $\pi$-tangle does not go to zero for higher values of acceleration. Due to relativistic effects the quantum discord 
of a system also does not go to zero in the infinite acceleration limit \cite{datta2009quantum}. For the tripartite 
entanglement, the non-vanishing nature is because the $\pi$-tangle is not a complete measure of entanglement. 
But for the quantum discord this is a fundamental feature of the underlying quantum correlations. From our work 
we find that there is a qualitative relationship between the total quantum correlations and the quantum coherence 
of a relativistic quantum system.  

In quantum information theory, it is not possible to distinguish between Bell states and the GHZ states through an estimation of entropy. But we can calculate the quantum coherence of these two states when they are in an inertial 
frame of reference. Here we again find that $C({|\psi_{Bell} \rangle})  = C({| \psi_{GHZ} \rangle})$ and so we will not be 
able to distinguish between these two states. This is because the GHZ state $\frac{1}{\sqrt{2}} (|000 \rangle + |111 \rangle)$ 
can be rewritten as an equivalent Bell state $\frac{1}{\sqrt{2}} (|0 \mathbf{0} \rangle + |1 \mathbf{1} \rangle)$ 
where $|{\bf 0} \rangle =|00 \rangle$ and $|{\bf 1} \rangle =|11 \rangle$ are redundantly encoded logical qubits 
encoding the same quantum information. Next we consider the situation where part of the system is moving with constant acceleration. When the single 
qubit in GHZ and Bell states move with constant acceleration it is not possible to distinguish between them. On 
the contrary when the logical qubit of the GHZ state is in an accelerated motion, then we have 
$C({|\psi_{Bell} \rangle}) \neq C({| \psi_{GHZ} \rangle})$ and the reason for this is that the size of the 
logical qubit under uniform acceleration is bigger than that of the regular qubit of the Bell state. This result is very interesting from a quantum information theory perspective where in general we cannot distinguish 
between Bell and GHZ states or between multipartite GHZ states with different number of qubits. But a measurement 
of quantum coherence combined with relativistic motion can differentiate them. Thus coherence measurement of 
accelerated systems can be used in quantum state discrimination.  

Next we consider a $N$-partite GHZ state and W-state and estimate the quantum coherence when $n < N$ qubits are accelerated. 
The GHZ and W-state represent the two extremes of coherence sharing. In the GHZ state, the coherence is present in a 
maximally entangled multipartite manner such that the loss of a single qubit destroys the entire coherence in the system. 
We find that in a GHZ state when the number of accelerating qubits $n$ increases, the coherence falls exponentially. 
The coherence in a W-state is shared in a local bipartite fashion. Here in the case of a W-states we observe that 
the coherence falls polynomially with increase in the number of accelerating qubits. In Fig.~\ref{fig10}, through a 
flow chart style we illustrate the quantum coherence decrease as we move from $n=2$ to $n>2$ quantum system.

The Bell states describe the maximally entangled states in a two qubit system.  When one of the qubits becomes non-inertial, 
then the coherence of the Bell state is given by $C_{\ell_{1}} =  Li_{-1/2} (\tanh^{2} r) / (\sinh^{2} r  \cosh r) $. There 
is only one way to entangle or correlate in a bipartite state. When we move to multipartite systems, there is more than one 
way to correlate the states. The GHZ and W-states denote the two extreme forms of correlation sharing where it is shared in 
a multipartite fashion in the former case and in a bipartite manner in the later one.  When more than one qubit is accelerated, 
the coherence falls exponentially in the GHZ state and polynomially in the W-state. As we move from the bipartite to the 
multipartite case, the linear fall of coherence observed in Bell type states changes to exponential fall for the GHZ states 
and polynomial decrease for the W-state. Hence the Bell type maximally entangled state clearly bifurcates into two extreme 
classes of entangled states with exponential fall (GHZ class) and polynomial fall (W-state) of coherence. This also proves 
that the W-state is more robust to Unruh decoherence compared to the GHZ state. This is because the coherence is shared in 
a genuinely multipartite manner in a GHZ state, but in a W-state it is shared only in a bipartite way. For large $N$ W-state, 
the loss of a few qubits results in a W-like state. This result might be useful for satellite based quantum communication, 
where some of the satellites are moving with very high velocities. In Ref.~\cite{wu2020quantum} the quantum coherence of 
multipartite W-states for Dirac field has been calculated in terms of the Kruskal modes in certain limiting cases. But 
in our work, we consider multipartite W-states of bosonic modes and compute quantum coherence in a very general setting 
where arbitrary number of qubits are being accelerated. In our method the Minkowski modes are converted into Rindler modes 
in the noninertial frame.

 A spherical non-rotating blackhole is described by a Schwarzchild space-time.  
 In the infinite acceleration 
limit, the Schwarzchild space-time can be approximately described by Rindler co-ordinates in Minkowski space time.  Hence our investigation in the infinite 
acceleration limit can be used to analyze quantum coherence in the context of black holes.  Our study shows that when Alice falls into a  black hole, 
she might still share quantum coherence with Bob and Charlie who are escaping from the black-hole.  
Hence we conclude that quantum correlations and quantum superposition present in a system do not completely vanish 
in the relativistic limit.

In our work we have used the single mode approximation where we use single frequency global modes. This can be experimentally realized using localized sources and detectors. 
Some proposals towards this end might include the use of localized projective measurements \cite{dragan2013localized}, homodyne detection \cite{downes2013quantum} 
and accelerated cavities \cite{downes2011entangling,friis2012kinematic}. Here we note that for a pure state, the $l_1$ norm measure of coherence is equal to 
the robustness of coherence \cite{wang2017directly}, which can be directly estimated using the interference of fringes. A possible method is the use of 
Berry phase atomic interferometry experiment \cite{rideout2012fundamental,martin2011using}, in which one of arm of the interferometer is under non-inertial motion. 
The phase difference between the inertial and non-inertial arms can give us the change in coherence due to the non-inertial motion. Hence an experimental verification 
of relativistic effects on quantum coherence might be an interesting future work. On the theoretical side an investigation on the quantum coherence effects in curved 
space time is also an interesting topic to explore.

\section*{Acknowledgements}
Chandrashekar Radhakrishnan was supported in part by a seed grant from 
IIT Madras to the Centre for Quantum Information, Communication and 
Computing. Peter Rohde is funded by an ARC Future Fellowship (project 
FT160100397). We are thankful to the anonymous reviewer for their comments on the single mode 
approximation, which was very useful. \\
\vspace{0.1cm}
\emph{Note: After submission of this manuscript we became aware of the work \cite{Wu_2021} which independently derived similar results.}

\bibliography{reference}

\end{document}